\documentclass[manuscript,screen,nonacm]{acmart}
\newcommand{\ket}[1]{\left|#1\right\rangle}

\newcommand{\braket}[1]{\left\langle#1\right\rangle}
\newcommand{\ketbra}[2]{\left|#1\right\rangle\!\left\langle#2\right|}

\theoremstyle{definition}

\theoremstyle{plain}
\newtheorem{proposition}{Proposition}

\usepackage{booktabs}
\usepackage{tabularx}
\usepackage{placeins}

\AtBeginDocument{}

\begin{document}
\raggedbottom

\title{Response-Threshold Framework for First-Peak Preservation in Noisy Grover Search}

\author{Ge Zhong}
\affiliation{%
  \institution{Faculty of Information Technology, Monash University}
  \city{Melbourne}
  \country{Australia}
}
\email{gzho0021@student.monash.edu}

\author{Charith Jayasekara}
\affiliation{%
  \institution{Faculty of Information Technology, Monash University}
  \city{Melbourne}
  \country{Australia}
}
\email{Charith.Jayasekara@monash.edu}

\author{Nisal De Silva}
\affiliation{%
  \institution{Faculty of Information Technology, Monash University}
  \city{Melbourne}
  \country{Australia}
}
\email{nisal.desilva1@monash.edu}

\author{Chetan Arora}
\affiliation{%
  \institution{Faculty of Information Technology, Monash University}
  \city{Melbourne}
  \country{Australia}
}
\email{chetan.arora@monash.edu}

\begin{abstract}
Grover search is executed only after integer iterations, so noise can alter the stopping decision even when a continuous first peak remains visible. We develop a response-threshold framework separating continuous peak motion, stroboscopic geometry, and the discrete sampled-success argmax. For a Drude--Lorentz environment, linearized HEOM supplies an a priori timing coefficient and sampled-probability differentials. The first-order layer is used as a conservative screen: well-separated points retain the nominal step, whereas near-threshold points escalate to finite-$\lambda$ evaluation. Across 260 unique HEOM points drawn from five overlapping validation sets, it detects all 15 observed stopping changes with zero regret on the changed subset; its residual errors are 45 one-step-early calls on otherwise stable points; the 245-point stable subset has mean regret $4.42\times10^{-3}$. An empirically validated two-stage workflow escalates every observed change while avoiding 160 of 260 unique full-HEOM propagations. Same-bath NZ2 reduces early calls. A heterogeneous three-qubit circuit exhibits three stopping-transition scales, each replicated across eight marked targets, and finite-shot recovery is $96.18\%$ overall, with failures localized to a near-degenerate boundary at $s=1.5$. A depth-refined transfer test supports the predicted local quadratic timing-to-loss relation.
\end{abstract}


\maketitle

\section{Introduction}
\label{sec:introduction}

Grover search and amplitude amplification provide a quadratic speed-up for unstructured search in the ideal oracle model~\cite{grover1996fast,boyer1998tight,brassard2002quantum,nielsen2010quantum}. Prior noisy-Grover work has largely focused on how noise modifies success probability or the conditions for retaining a speed-up. Perturbative analyses quantify decoherence corrections to Grover dynamics~\cite{azuma2002decoherence,azuma2005higher}; faulty-input and faulty-oracle studies establish algorithmic limits under imperfect queries~\cite{hoyer2003quantum,regev2008impossibility}; and error-correction, error-suppression, mitigation, and hardware-oriented studies ask how much search performance can be recovered or retained under realistic noise~\cite{botsinis2016quantum,pokharel2024better,abughanem2025characterizing,srivastava2025switches}. Broader NISQ and error-mitigation reviews provide the surrounding implementation context~\cite{bharti2022noisy,cai2023quantum}. These lines of work motivate noise-aware search, but they do not by themselves determine whether a visible noisy first peak selects the same integer Grover iteration as the ideal trajectory.

This paper isolates that stopping layer. We separate three objects: the continuous first-peak time, the geometric location of that peak inside the ideal stroboscopic cell, and the actual discrete maximizer of the sampled probabilities. The continuous-cell geometry is a useful warning signal, while for a general asymmetric trajectory sampled reordering is determined directly by the sampled probabilities; a midpoint rule follows only under additional local symmetry. The resulting framework therefore combines a signed timing response with direct sampled-probability response differentials.

The closest algorithmic comparison is the noise-aware stopping-step prediction of Leng et al.~\cite{leng2025noise}, whose native objective minimizes a restart-aware runtime constructed from a noise-informed success predictor. Our primary objective is different: first-peak stopping preservation on the local sampled candidate set. We therefore compare the two methods only under their explicitly stated objectives rather than collapsing them into a single method ranking. The gap addressed here is not another mitigation scheme or a full-dynamics surrogate, but a stopping-layer theory that predicts where structured noise can threaten the integer sampled argmax and identifies when a low-order response is no longer sufficient. Operationally, the workflow is screening then escalation: use one a priori linearized response to locate stable and near-threshold regimes, and invoke finite-$\lambda$ dynamics only where the sampled ordering approaches a predicted transition.

The main contributions are:
\begin{enumerate}
    \item \textbf{A priori first-peak response.} For the Drude--Lorentz validation model, the first-order success response $\chi_q^{\mathrm{lin}}$ and signed timing coefficient $G_q^{\mathrm{lin}}$ are computed from a linearized HEOM driven only by the ideal trajectory, bath parameters, and coupling geometry. Finite-$\lambda$ HEOM peak shifts are used only afterwards as independent validation data.

    \item \textbf{Sampled stopping diagnostics.} The reduced dynamics is calibrated to one full Grover iteration through $\Omega_G\tau_G=4\theta$. The continuous-cell threshold uses the direction-aware stroboscopic margin, while the operational sampled-step criterion compares the ideal gaps $\Delta_{0,j}$ directly with the response differentials $D_{q,j}$. The three-point local candidate set is validated against the complete first Grover lobe, giving identical HEOM argmax decisions on all 260 unique tested points.

    \item \textbf{Threshold validity, escalation, and bath sensitivity.} Deterministic numerical uncertainty is propagated into uncertainty-separated first-order indicator intervals. Broad HEOM domains establish a no-crossing regime, while targeted scans expose both genuine crossings and higher-order-stabilized controls. At $\eta_{\mathrm{esc}}=0.75$, the two-stage workflow escalates all 15 observed changes on the 260 unique physical HEOM points while reducing unique full-HEOM calls from 260 to 100. Bath variation is separated into recalibrated threshold-topology robustness and fixed-operating-$\lambda$ mismatch; same-bath NZ2 is used as a finite-$\lambda$ escalation comparator rather than as a replacement for the first-order transition scale.

    \item \textbf{Operational validation and local transfer structure.} A heterogeneous device-inspired three-qubit circuit provides an independent sampled-stopping test beyond the reduced HEOM model: the exact stopping transition occurs at three noise scales, each replicated across all eight marked targets, and the first-order rule recovers all three transition scales while its finite-shot failures are localized near the narrow-gap boundary. A depth-refined continuous-time test additionally supports the predicted \emph{local} quadratic relation between cross-response mistiming and transfer loss.
\end{enumerate}

\section{Sampled Stopping Geometry and Structured-Noise Realization}
\label{sec:geometry-model}

\subsection{Grover-active stopping layer}
\label{subsec:grover-active-subspace}

For a single marked target $\ket{t}$, let $\ket{r}$ denote the normalized superposition of unmarked states. The ideal Grover dynamics lies in
\[
    \mathcal{H}_G=\operatorname{span}\{\ket{t},\ket{r}\},
\]
with initial state
\[
    \ket{\psi_0}=\sin\theta\,\ket{t}+\cos\theta\,\ket{r},
    \qquad \sin\theta=\frac{1}{\sqrt N}.
\]
We use the continuous reduced representation
\begin{equation}
    H_S=-\frac{\Omega_G}{2}\sigma_y,
    \label{eq:reduced-grover-hamiltonian}
\end{equation}
where $\Omega_G$ sets the continuous rotation time scale in $\mathcal H_G$. The target population is
\[
    P_{\mathrm{succ}}(t)=\braket{t|\rho_S(t)|t}.
\]
We write $P_0(t;\theta)$ for the ideal trajectory and $P_q(t)$ for a noisy response labelled by $q$; the concrete coupling labels $q\in\{z,x\}$ are specified in Section~\ref{subsec:structured-noise-main}. The reduced trajectory is mapped to complete Grover iterations by the standard sampling calibration
\begin{equation}
    \Omega_G\tau_G=4\theta.
    \label{eq:stroboscopic-calibration}
\end{equation}
Thus one integer sampled index $j$ represents one complete Grover iteration. In the ideal reduced dynamics,
\begin{equation}
    P_0(j\tau_G;\theta)=\sin^2\!\bigl[(2j+1)\theta\bigr],
    \qquad j\in\mathbb Z_{\ge0},
    \label{eq:standard-grover-sampled-probability}
\end{equation}
which is the standard single-target Grover success law. The circuit-level test in Section~\ref{subsec:circuit-validation} uses the same integer-iteration convention directly at the gate level.

Let $t_0^*(\theta)$ be the first continuous maximum of the ideal trajectory. Its location in standard-iteration units is
\begin{equation}
    j_{\mathrm{cont}}(\theta)
    =\frac{t_0^*(\theta)}{\tau_G(\theta)}
    =\frac{\pi}{4\theta}-\frac12.
    \label{eq:continuous-grover-index}
\end{equation}
To separate the ideal first Grover lobe from the local stopping comparison, define the complete nonnegative first-lobe index set
\begin{equation}
    \mathcal J_{\mathrm{lobe}}(\theta)
    =\left\{j\in\mathbb Z_{\ge0}:(2j+1)\theta\le\pi\right\}.
    \label{eq:first-lobe-index-set}
\end{equation}
The ideal sampled winner is fixed before any noisy trajectory is examined,
\begin{equation}
    j_0^*(\theta)
    =\operatorname*{arg\,max}_{j\in\mathcal J_{\mathrm{lobe}}(\theta)}
    P_0(j\tau_G;\theta),
    \label{eq:ideal-sampled-winner}
\end{equation}
with exact ties resolved by choosing the smaller iteration index. The operational local candidate set is then the symmetric three-point neighbourhood of that ideal sampled winner,
\begin{equation}
    \mathcal J_3(\theta)
    =\{j_0^*-1,j_0^*,j_0^*+1\}
    \cap\mathcal J_{\mathrm{lobe}}(\theta).
    \label{eq:standard-candidate-set}
\end{equation}
For every $N\ge4$, one has $\theta\le\pi/6$, so $\mathcal J_{\mathrm{lobe}}$ contains at least $\{0,1,2\}$ and the ideal first-lobe winner satisfies $j_0^*\ge1$; hence $\mathcal J_3$ cannot collapse to a singleton. In the limiting $N=4$ case, $\theta=\pi/6$, $j_0^*=1$, and
$\mathcal J_3=\mathcal J_{\mathrm{lobe}}=\{0,1,2\}$.
The noisy sampled stopping iteration used by the local stopping layer is
\begin{equation}
    j_q^*=\operatorname*{arg\,max}_{j\in\mathcal J_3(\theta)}P_q(j\tau_G),
    \label{eq:sampled-stopping-step}
\end{equation}
using the same smaller-index tie rule. Thus $j_0^*$ is defined against the complete ideal first lobe, while the noisy stopping comparison is performed on the pre-specified local neighbourhood $\mathcal J_3$. Continuous peak motion remains diagnostic; sampled probability ordering decides the implemented integer Grover iteration.

\begin{figure}[t]
    \centering
    \includegraphics[width=0.98\linewidth]{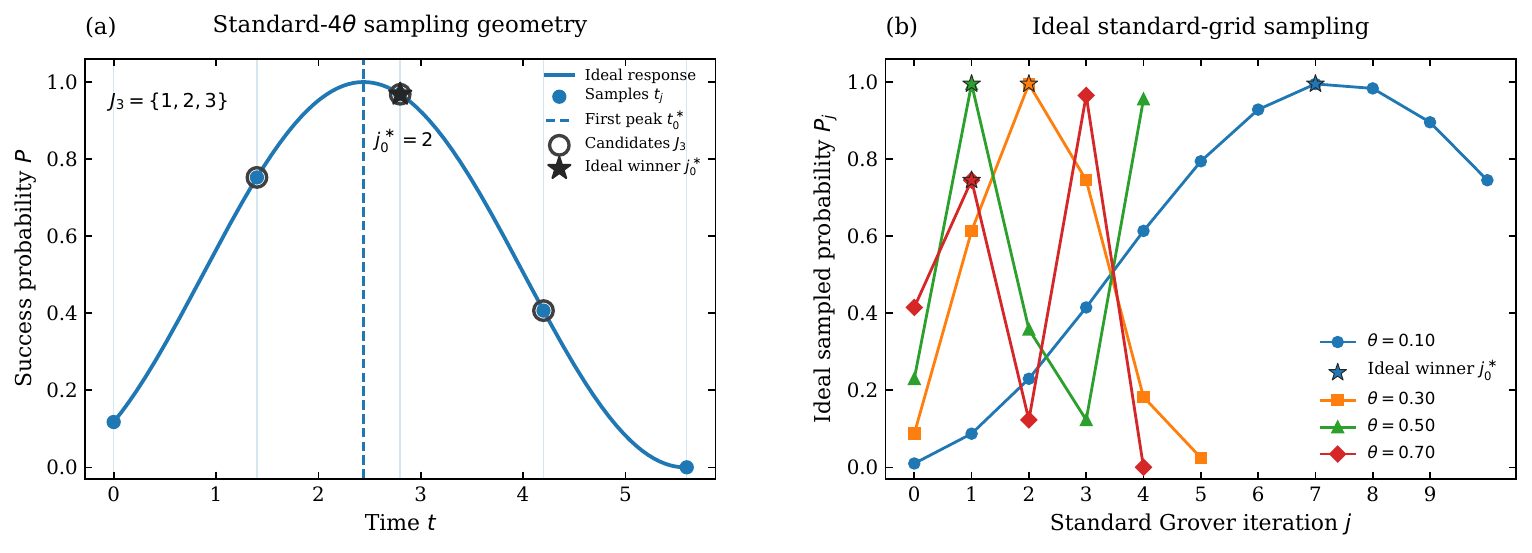}
    \caption{Standard-$4\theta$ sampling geometry. Panel (a) shows a representative continuous ideal response, the standard sampling times $t_j=j\tau_G$, the first continuous peak $t_0^*$, the three-point candidate neighbourhood $\mathcal J_3=\{j_0^*-1,j_0^*,j_0^*+1\}\cap\mathcal J_{\mathrm{lobe}}$, and the ideal sampled winner $j_0^*$. Panel (b) shows ideal sampled probabilities for representative Grover angles under $P_0(j\tau_G)=\sin^2[(2j+1)\theta]$. Throughout, $\tau_G=4\theta/\Omega_G$, so each integer $j$ denotes one complete Grover iteration.}
    \Description{Two-panel figure illustrating the standard four-theta sampling geometry and ideal sampled Grover probabilities at integer iteration indices.}
    \label{fig:sampling-geometry}
\end{figure}

\subsection{Structured bosonic response}
\label{subsec:structured-noise-main}

The structured validation model couples Eq.~\eqref{eq:reduced-grover-hamiltonian} to a Drude--Lorentz bosonic bath with
\begin{equation}
    J(\omega)=\frac{2\lambda\gamma\omega}{\omega^2+\gamma^2},
    \label{eq:drude-lorentz-spectral-density}
\end{equation}
where $\lambda$ is the coupling scale, $\gamma^{-1}$ is the memory scale, and $T$ is the bath temperature~\cite{ishizaki2005quantum,shi2009efficient,tanimura1989time,tanimura2020numerically,weiss2012quantum}. We compare two coupling geometries,
\[
    S_q\in\{\sigma_z,\sigma_x\},\qquad q\in\{z,x\},
\]
under the same Hamiltonian, initial state, bath spectrum, and sampled stopping rule.

HEOM has two distinct roles. Full HEOM provides finite-memory validation trajectories. Separately, the weak-coupling hierarchy is linearized about the ideal Grover trajectory to obtain an a priori first-order response before any finite-$\lambda$ peak shifts are examined. The detailed hierarchy and its linearization are given in Appendix~\ref{app:heom-details}. Quantitative convergence fixes the production hierarchy depth at $L_{\mathrm{prod}}=3$ and the retained Matsubara order at $N_{k,\mathrm{prod}}=6$, as documented in Appendix~\ref{app:numerical-protocols}.

\subsection{Reference model and scope}
\label{subsec:reference-models-main}

The model-comparison layer uses a same-bath second-order Nakajima--Zwanzig model (NZ2) built from the same $H_S$, $S_q$, Drude--Lorentz correlation function, and $(\lambda,\gamma,T)$ as HEOM, without fitting an HEOM damping rate, phase shift, or peak time. This comparator tests which stopping-threshold features survive a second-order reduced description. Its explicit memory-kernel equation is given in Appendix~\ref{app:reference-model-details}.

The two-dimensional model is a stopping-layer abstraction, not a complete hardware model. Hardware-level gate and measurement noise can alter the effective target-population trajectory~\cite{abughanem2025characterizing}. Error-correction-based treatments provide the broader fault-tolerance context~\cite{botsinis2016quantum,fowler2012surface,terhal2015quantum}. Section~\ref{subsec:circuit-validation} therefore adds a separate small multi-qubit circuit test of the discrete sampled-response layer rather than treating the reduced HEOM model as a full-circuit simulator.

\section{Predictive Response-Threshold Framework}
\label{sec:predictive-framework}

\subsection{A priori timing response}
\label{subsec:apriori-response}

In a local first-peak neighbourhood $I$ containing $t_0^*$, suppose $P_q(t;\lambda,\theta)$ is jointly $C^2$ in $(t,\lambda)$ for sufficiently small $|\lambda|$ and admits the first-order expansion
\begin{equation}
    P_q(t;\lambda,\theta)
    =P_0(t;\theta)+\lambda\chi_q(t;\gamma,T,\theta)+\mathcal E_q(t,\lambda;\theta),
    \label{eq:success-weak-coupling}
\end{equation}
where $P_0\in C^3(I)$, $\chi_q\in C^2(I)$, and the remainder has uniform derivative control
\[
    \sup_{t\in I}\!\left(|\partial_t\mathcal E_q(t,\lambda;\theta)|+|\partial_t^2\mathcal E_q(t,\lambda;\theta)|\right)
    \le C_I\lambda^2
\]
for sufficiently small $\lambda$. Let the ideal first maximum be isolated and non-degenerate,
$P_0'(t_0^*;\theta)=0$ and
$\kappa_0(\theta)=-P_0''(t_0^*;\theta)>0$.

\begin{proposition}[First-order first-peak response]
\label{prop:first-order-first-peak-response}
Under the preceding local regularity, uniform-remainder, and non-degeneracy assumptions, there exists a neighbourhood $I_0\subset I$ of $t_0^*$ such that, for sufficiently small $\lambda$, a unique stationary maximum $t_q^*(\lambda;\theta)\in I_0$ is continuously connected to $t_0^*(\theta)$, and
\begin{equation}
    t_q^*(\lambda;\theta)
    =t_0^*(\theta)+\lambda G_q(\gamma,T;\theta)+O(\lambda^2),
    \label{eq:first-peak-expansion}
\end{equation}
with
\begin{equation}
    G_q(\gamma,T;\theta)
    =\frac{\dot\chi_q(t_0^*;\gamma,T,\theta)}{\kappa_0(\theta)}.
    \label{eq:Gq-definition}
\end{equation}
\end{proposition}

For the HEOM realization, $\chi_q^{\mathrm{lin}}$ is obtained by solving the linearized hierarchy from the ideal trajectory and unit-coupling bath coefficients. The model-based coefficient and first-order timing prediction are
\begin{equation}
    G_q^{\mathrm{lin}}(\gamma,T;\theta)
    =\frac{\dot\chi_q^{\mathrm{lin}}(t_0^*;\gamma,T,\theta)}{\kappa_0(\theta)},
    \qquad
    t_{q,\mathrm{pred}}^{*(1)}=t_0^*+\lambda G_q^{\mathrm{lin}}.
    \label{eq:Gq-linearized-prediction}
\end{equation}
The empirical coefficient $\widehat G_q^{\mathrm{traj}}$, extracted from finite-$\lambda$ full-HEOM peak times, is used only as an independent comparator. The proof of Proposition~\ref{prop:first-order-first-peak-response} and the HEOM response equations are in Appendices~\ref{app:proofs} and~\ref{app:heom-details}.

For comparisons between the two coupling geometries, define the channel-splitting coefficient
\begin{equation}
    G_{xz}(\gamma,T;\theta)=G_x(\gamma,T;\theta)-G_z(\gamma,T;\theta).
    \label{eq:channel-splitting-response}
\end{equation}
Equation~\eqref{eq:first-peak-expansion} then gives the first-order timing split
\begin{equation}
    \Delta t_{xz}^*
    \equiv t_x^*-t_z^*
    =\lambda G_{xz}+O(\lambda^2).
    \label{eq:cross-response-timing-split}
\end{equation}

\subsection{Continuous-cell and sampled-step diagnostics}
\label{subsec:stopping-diagnostics}

The ideal first peak lies inside the stroboscopic cell centred at $j_0^*\tau_G$. Its one-sided margins are
\begin{equation}
\begin{aligned}
    m_0^-&=t_0^*-\left(j_0^*-\frac12\right)\tau_G,\\
    m_0^+&=\left(j_0^*+\frac12\right)\tau_G-t_0^*.
\end{aligned}
\label{eq:stroboscopic-margins}
\end{equation}
For $G_q\neq0$, let $m_{0,q}^{\mathrm{dir}}=m_0^+$ if $G_q>0$ and $m_0^-$ if $G_q<0$. The continuous-cell threshold and indicator are
\begin{equation}
    \lambda_{\mathrm{cell},q}^{(1)}
    =\frac{m_{0,q}^{\mathrm{dir}}}{|G_q|},
    \qquad
    \eta_{\mathrm{cell},q}^{(1)}
    =\frac{\lambda|G_q|}{m_{0,q}^{\mathrm{dir}}}.
    \label{eq:continuous-cell-threshold}
\end{equation}
These are geometric diagnostics only; they do not certify the sampled argmax for a general asymmetric trajectory.

For each competitor $j\neq j_0^*$ in $\mathcal J_3$, define
\begin{equation}
    \Delta_{0,j}=P_0(j_0^*\tau_G)-P_0(j\tau_G)>0,
    \qquad
    D_{q,j}=\chi_q(j\tau_G)-\chi_q(j_0^*\tau_G).
    \label{eq:ideal-sampled-gap}
\end{equation}
Here $\chi_q$, $G_q$, and $D_{q,j}$ denote the generic first-order response objects used in the formal statements. In the HEOM realization they are instantiated by
\begin{equation}
    D_{q,j}^{\mathrm{lin}}
    =
    \chi_q^{\mathrm{lin}}(j\tau_G)
    -
    \chi_q^{\mathrm{lin}}(j_0^*\tau_G),
    \qquad
    (\chi_q,G_q,D_{q,j})
    \mapsto
    (\chi_q^{\mathrm{lin}},G_q^{\mathrm{lin}},D_{q,j}^{\mathrm{lin}}),
    \label{eq:heom-response-realization-map}
\end{equation}
and these a priori linearized-HEOM quantities are the inputs to the numerical predictor and uncertainty analysis. The corresponding first-order discrete threshold is
\begin{equation}
    \lambda_{\mathrm{step},q}^{(1)}
    =
    \min_{\substack{j\in\mathcal J_3,\;j\neq j_0^*\\D_{q,j}>0}}
    \frac{\Delta_{0,j}}{D_{q,j}},
    \label{eq:first-order-sampled-step-threshold}
\end{equation}
with $\lambda_{\mathrm{step},q}^{(1)}=\infty$ if no competitor has $D_{q,j}>0$. The associated dimensionless indicator is
\begin{equation}
    \eta_{\mathrm{step},q}^{(1)}
    =\max_{\substack{j\in\mathcal J_3,\;j\neq j_0^*}}
    \frac{\lambda[D_{q,j}]_+}{\Delta_{0,j}},
    \qquad [x]_+=\max\{x,0\}.
    \label{eq:first-order-boundary-risk-indicator}
\end{equation}
In the HEOM realization, the same first-order response gives the explicit point predictor
\begin{equation}
    P_{q,\mathrm{pred}}^{(1)}(j)
    =P_0(j\tau_G)+\lambda\chi_q^{\mathrm{lin}}(j\tau_G),
    \qquad
    j_{q,\mathrm{pred}}^{*(1)}
    =\operatorname*{arg\,max}_{j\in\mathcal J_3}P_{q,\mathrm{pred}}^{(1)}(j).
    \label{eq:first-order-predicted-sampled-argmax}
\end{equation}

\begin{proposition}[First-order sampled-argmax criterion]
\label{prop:first-order-sampled-cell}
If Eq.~\eqref{eq:success-weak-coupling} holds uniformly on $\mathcal J_3$ and $j_0^*$ is the unique ideal sampled maximizer, write
\begin{equation}
    P_q(j_0^*\tau_G)-P_q(j\tau_G)
    =\Delta_{0,j}-\lambda D_{q,j}+\mathcal E^{\mathrm{step}}_{q,j}(\lambda),
    \label{eq:sampled-gap-expansion}
\end{equation}
with a uniform local bound $|\mathcal E^{\mathrm{step}}_{q,j}(\lambda)|\le C_q\lambda^2$ for all $j\in\mathcal J_3$ and $0\le\lambda\le\bar\lambda$. Then
\[
    \Delta_{0,j}-\lambda D_{q,j}>C_q\lambda^2
    \quad\text{for every }j\neq j_0^*
\]
is sufficient for preservation, $j_q^*=j_0^*$. Conversely, if some competitor satisfies
\[
    \lambda D_{q,j}-\Delta_{0,j}>C_q\lambda^2,
\]
then that competitor outranks $j_0^*$ and hence $j_q^*\neq j_0^*$. In particular, for any fixed $0<\delta<1$, $\eta_{\mathrm{step},q}^{(1)}\le1-\delta$ is sufficient whenever $C_q\lambda^2<\delta\min_{j\neq j_0^*}\Delta_{0,j}$; the analogous reversal conclusion follows when some $\lambda D_{q,j}/\Delta_{0,j}\ge1+\delta$ and $C_q\lambda^2<\delta\Delta_{0,j}$ for that competitor.
\end{proposition}

Thus a predicted integer iteration and a certificate are not the same object. Equation~\eqref{eq:first-order-predicted-sampled-argmax} returns a candidate argmax; Proposition~\ref{prop:first-order-sampled-cell} states the explicit remainder control under which the first-order ordering is safely separated from the transition.

\paragraph{Scope of the first-order layer.}
The first-order construction is intended as a screening layer rather than an order-by-order reconstruction of the full HEOM dynamics. At fixed bath parameters, one linearized solve provides the sampled response and transition scale across coupling strength; operating points approaching that scale can then be escalated to direct finite-$\lambda$ evaluation. An explicit second-order response could refine this screening layer, but is not required for the present screening-and-escalation workflow.

\subsection{Uncertainty-aware screening bounds}
\label{subsec:uncertainty-main}

Let the reported inputs satisfy
\[
    |\widehat G_q-G_q|\le\epsilon_G,
    \quad
    |\widehat D_{q,j}-D_{q,j}|\le\epsilon_{D,j},
    \quad
    |\widehat\Delta_{0,j}-\Delta_{0,j}|\le\epsilon_{\Delta,j},
\]
with analogous one-sided margin bounds. For the discrete criterion, provided $\widehat\Delta_{0,j}>\epsilon_{\Delta,j}$, define
\begin{equation}
\begin{aligned}
    \eta_{\mathrm{step},q}^{-}
    &=\max_{j\neq j_0^*}
    \frac{\lambda[\widehat D_{q,j}-\epsilon_{D,j}]_+}
    {\widehat\Delta_{0,j}+\epsilon_{\Delta,j}},\\
    \eta_{\mathrm{step},q}^{+}
    &=\max_{j\neq j_0^*}
    \frac{\lambda[\widehat D_{q,j}+\epsilon_{D,j}]_+}
    {\widehat\Delta_{0,j}-\epsilon_{\Delta,j}}.
\end{aligned}
\label{eq:uncertain-sampled-step-indicator-bounds}
\end{equation}
Then $\eta_{\mathrm{step},q}^{+}\le1-\delta$ gives an uncertainty-separated first-order same-step condition, whereas $\eta_{\mathrm{step},q}^{-}\ge1+\delta$ gives an uncertainty-separated first-order reversal condition. These become sufficient certificates for the full sampled ordering only when the explicit remainder bound in Proposition~\ref{prop:first-order-sampled-cell} is also satisfied. If neither first-order inequality is separated, the point is reported as uncertainty-sensitive. These intervals quantify numerical uncertainty in the response and sampled quantities; sensitivity to the physical inputs $(\gamma,T,\lambda)$ is assessed separately rather than folded into $\epsilon_G$, $\epsilon_{D,j}$, or $\epsilon_{\Delta,j}$. The continuous-cell error bounds, physical-input sensitivity check, and higher-order diagnostics are given in Appendix~\ref{app:uncertainty-details}.

\subsection{Local transfer loss and conditional large-instance interpretation}
\label{subsec:transfer-main}

For two responses $c$ and $d$, define
\begin{equation}
    R_{c\leftarrow d}=P_c(t_c^*)-P_c(t_d^*).
    \label{eq:transfer-loss-definition}
\end{equation}
If $P_c$ has a smooth non-degenerate local maximum with curvature $\kappa_c=-P_c''(t_c^*)>0$, then
\begin{proposition}[Quadratic local stopping-transfer loss]
\label{prop:quadratic-transfer-loss}
\begin{equation}
    R_{c\leftarrow d}
    =\frac12\kappa_c(t_d^*-t_c^*)^2
    +O(|t_d^*-t_c^*|^3).
    \label{eq:quadratic-transfer-loss}
\end{equation}
\end{proposition}
A first-order timing shift can therefore be stopping-relevant even when the local success loss remains second order. Combining Eq.~\eqref{eq:cross-response-timing-split} with Proposition~\ref{prop:quadratic-transfer-loss} gives, for transfer between the $x$ and $z$ responses,
\begin{equation}
    R_{c\leftarrow d}
    =\frac12\kappa_c\lambda^2G_{xz}^2+O(\lambda^3),
    \qquad c,d\in\{x,z\},\ c\neq d.
    \label{eq:cross-response-loss-scaling}
\end{equation}

For the large-instance interpretation, we impose the non-resonance qualification stated below. If $\theta(N)\sim N^{-1/2}$, the ideal first peak stays a fixed fractional distance from half-integer stroboscopic boundaries, and $|G_q|=\Theta(\theta^\alpha)$ for a response-scaling exponent $\alpha$, then
\begin{equation}
    \lambda_{\mathrm{cell},q}^{(1)}
    =\Theta(\theta^{1-\alpha})
    =\Theta\!\left(N^{-(1-\alpha)/2}\right).
    \label{eq:formal-continuous-cell-threshold-scaling}
\end{equation}
This scaling is conditional on the non-resonant instance sequence
specified in Appendix~\ref{app:large-instance-details} and is not
asserted uniformly over all $N$. Under the corresponding
cell-preserving scaling and uniform second-order control, the local
cross-response transfer loss is $O(N^{-1})$. The arithmetic non-resonance condition is detailed in Appendix~\ref{app:large-instance-details}, the response-aware calibration bound in Appendix~\ref{app:calibration-bound}, and the uncertainty and higher-order expansions in Appendix~\ref{app:uncertainty-details}.

\section{Numerical Validation of Stability and Threshold Crossing}
\label{sec:numerical-validation}

\subsection{Numerical protocol, candidate-set validation, and a priori response}
\label{subsec:numerical-protocol-main}
\label{subsec:apriori-validation}

All full-HEOM production scans use $L_{\mathrm{prod}}=3$ and $N_{k,\mathrm{prod}}=6$, with the standard operational sampling rule $\Omega_G\tau_G=4\theta$ from Section~\ref{sec:geometry-model}. The primary domain covers $\theta=0.30,0.35,\ldots,0.70$ and $\lambda=0.005,0.010,\ldots,0.030$ for both $S_z$ and $S_x$, giving 108 channel-points at $\gamma=0.5$ and $T=1.0$. The coupling-hold-out test is a pre-specified $\lambda$ split within this same $(\theta,q,\gamma,T)$ parameter grid: the low-coupling values are used for the empirical timing calibration and the higher-coupling values are reserved for evaluation. It is therefore a coupling hold-out, not an independent external domain. The small-angle domain uses $\theta\in\{0.10,0.125,0.15,0.20,0.25,0.30,0.35,0.40\}$ with $\lambda\in\{0.0025,0.0050,0.0075,0.0100\}$ for both channels. All sampled quantities in this section are evaluated on the grid generated by Eq.~\eqref{eq:stroboscopic-calibration}.

The continuous a priori response layer is validated independently of the discrete sampling layer. The linearized solver passes the $\lambda\to0$ finite-difference check and preserves trace and Hermiticity at machine precision. Across nine angles and two coupling geometries, all 18 coefficients $G_q^{\mathrm{lin}}$ have the correct sign relative to independently fitted finite-coupling HEOM slopes. The pooled coefficient MAE is $0.014438$, the mean relative error is $0.2303\%$, and the maximum relative error is $0.3808\%$. Reusing the same $G_q^{\mathrm{lin}}$ without finite-$\lambda$ refitting gives a pooled first-peak timing MAE of $0.001822$ on the 72-point weak-coupling validation set. Detailed convergence, extraction, and physicality protocols are collected in Appendix~\ref{app:numerical-protocols}.

Before comparing stopping rules, we compare the local three-point candidate geometry with the complete nonnegative first Grover lobe. All five pre-specified validation datasets were reevaluated under $\mathcal J_3$: together they contain 346 dataset instances corresponding to 260 unique physical HEOM points after deduplication. On all 260 unique points, the HEOM argmax over $\mathcal J_3$ agrees with the argmax over $\mathcal J_{\mathrm{lobe}}$ exactly. Expanding the previous local comparison to $\mathcal J_3$ changes zero HEOM stopping steps, zero HEOM stable/change labels, zero first-order predicted steps, and zero first-order stable/change labels. The local rule used below is therefore numerically complete on the tested domain while retaining the same substantive stopping outcomes.

\subsection{Broad-domain stopping stability}
\label{subsec:broad-stability-results}
\label{subsec:full-scan-results}
\label{subsec:heldout-results}

\begin{figure}[t]
    \centering
    \includegraphics[width=0.98\linewidth]{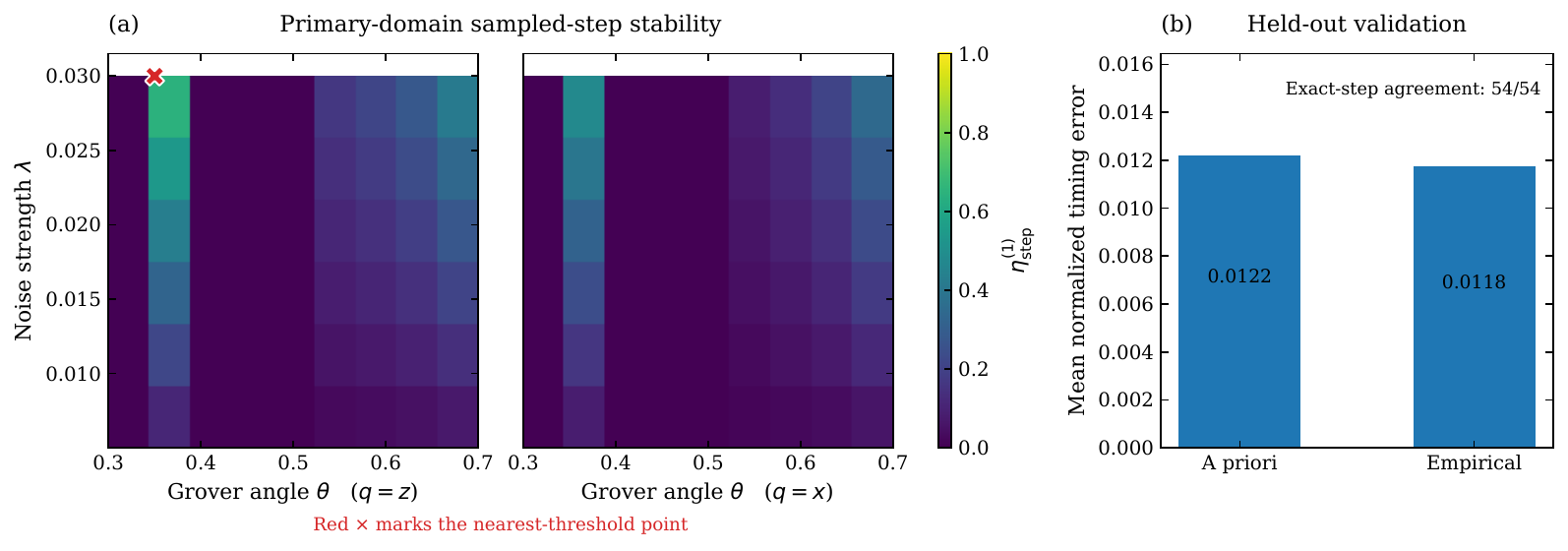}
    \caption{Broad-domain stopping stability. Panel (a) shows the first-order stopping indicator $\eta_{\mathrm{step}}^{(1)}$ over the 108-point primary scan for the two coupling channels; the marked point is the closest primary point to the reversal threshold. Panel (b) gives the normalized errors on the pre-specified $\lambda$-hold-out evaluation set, for which exact sampled-step agreement is $54/54$.}
    \Description{Two-panel validation figure showing primary-domain stopping indicators and normalized timing errors on the coupling hold-out split.}
    \label{fig:broad-stability}
\end{figure}

Figure~\ref{fig:broad-stability} summarizes the broad-domain result on the sampled Grover grid. On the 108-point primary scan, the first-order sampled predictor, the full-HEOM sampled optimum, and the ideal standard-Grover optimum agree at every point: exact-step agreement is $108/108$ and the mean absolute step error is zero.
No point exhibits a predicted or observed integer-step change. All 108 points are nominal preservation and are also uncertainty-separated on the preservation side. The closest production point to the sampled reversal threshold is $(\theta,\lambda,q)=(0.35,0.030,z)$, where $\eta_{\mathrm{step}}^{(1)}=0.640713$ and the deterministic interval is approximately $[0.640423,0.641003]$. The smallest first-order reversal threshold in the primary domain is $\lambda_{\mathrm{step}}^{(1)}=0.046823$, safely above the largest scanned coupling $\lambda=0.030$. Thus the primary domain is a robust no-crossing regime.

The $\lambda$-hold-out test leads to the same operational conclusion. Calibration uses $\lambda=0.005,0.010,0.015$ and evaluation reserves $0.020,0.025,0.030$ on the same angle, channel, and bath grid. Continuous timing errors remain about $1.2\%$ in normalized iteration units, while the sampled layer gives exact predictor--HEOM agreement at all 54 coupling-hold-out points, with $j_{q,\mathrm{pred}}^{*(1)}=j_q^*=j_0^*$ throughout.

\begin{figure}[t]
    \centering
    \includegraphics[width=0.64\linewidth]{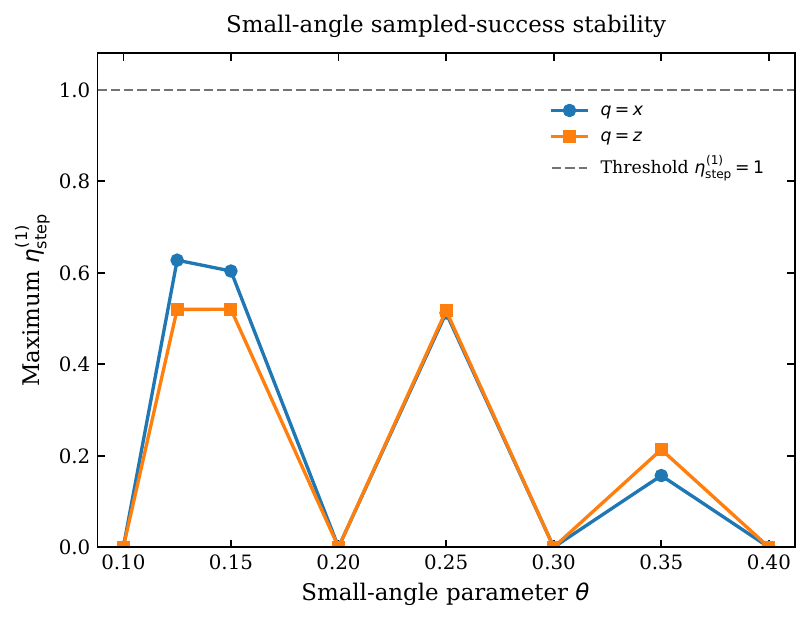}
    \caption{Small-angle sampled-success stability. The plotted envelope shows the maximum first-order stopping indicator $\eta_{\mathrm{step}}^{(1)}$ at each tested angle for $q=x$ and $q=z$. All 64 points remain below the nominal reversal threshold $\eta_{\mathrm{step}}^{(1)}=1$, and both the fixed ideal and first-order rules recover the HEOM sampled optimum at $64/64$ points.}
    \Description{Single-panel small-angle benchmark showing the sampled-success stopping-indicator envelope for the two coupling channels.}
    \label{fig:small-angle-leng}
\end{figure}

The 64-point small-angle benchmark is also stable on this grid. Both the fixed ideal rule and the first-order predictor agree with the HEOM sampled optimum at $64/64$ points, and no sampled-step change occurs. The maximum first-order stopping indicator is $\eta_{\mathrm{step}}^{(1)}=0.627565$, attained at $(\theta,\lambda,q)=(0.125,0.010,x)$, leaving a distance $0.372435$ to the first-order reversal threshold. The adapted Leng implementation also reproduces the 64-point native-runtime benchmark, but because no Leng step transition occurs on this domain we treat that result only as an implementation-consistency check and defer its protocol to Appendix~\ref{app:small-angle-protocol}. Figure~\ref{fig:small-angle-leng} therefore reports only the sampled-success stability envelope. Hence these three broad weak-to-moderate-noise domains consistently support stopping stability away from the predicted boundary.

\subsection{Theory-guided threshold crossings}
\label{subsec:threshold-crossing-results}

The broad scans above remain below the first-order stopping thresholds. To test whether the threshold theory identifies genuine Grover-step changes, we therefore construct a separate pre-specified stress test. Four candidate families are selected before the additional HEOM evaluation by taking, within the primary and small-angle domains and separately for $q\in\{z,x\}$, the smallest finite $\lambda_{\mathrm{step}}^{(1)}$. For each candidate we evaluate the five fixed ratios
\[
\lambda/\lambda_{\mathrm{step}}^{(1)}\in\{0.75,0.90,1.00,1.10,1.25\},
\]
giving 20 full-HEOM points with no post-hoc extension of the scan.

\begin{figure}[t]
    \centering
    \includegraphics[width=0.98\linewidth]{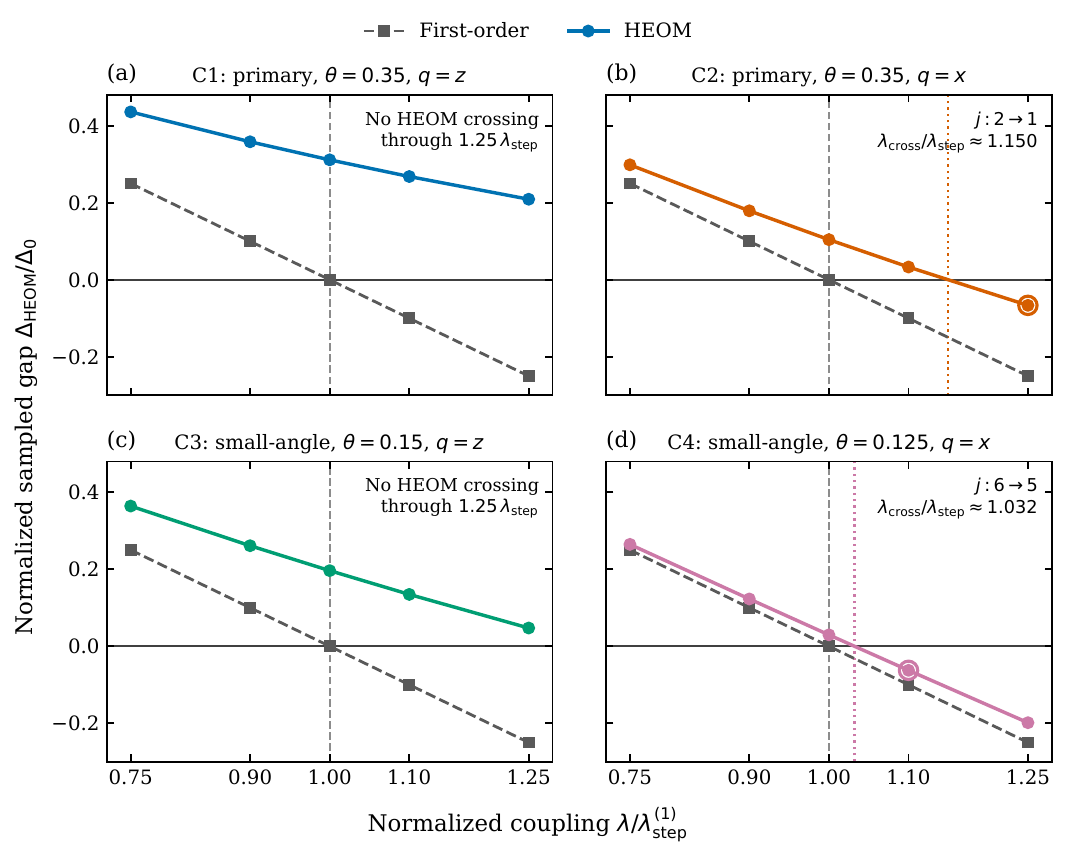}
    \caption{Theory-guided threshold-crossing validation. The four candidate families are sampled at fixed ratios $\lambda/\lambda_{\mathrm{step}}^{(1)}$. C2 and C4 exhibit genuine full-HEOM stopping transitions, whereas C1 and C3 remain non-crossing through $1.25\lambda_{\mathrm{step}}^{(1)}$. The annotated crossing ratios are coarse stress-scan estimates; hierarchy-refined roots are reported in Section~\ref{subsec:threshold-refinement-results}. The dashed first-order curves identify the nominal transition scale before the finite-coupling HEOM evaluation.}
    \Description{Four-panel plot of normalized sampled gaps versus normalized coupling for two crossing candidates and two non-crossing controls, comparing first-order predictions with full HEOM.}
    \label{fig:threshold-crossings}
\end{figure}

Figure~\ref{fig:threshold-crossings} shows the resulting sampled-gap trajectories. The two $x$-coupled candidates undergo genuine full-HEOM crossings. For C2 $(\theta=0.35,q=x)$, first order predicts $\lambda_{\mathrm{step}}^{(1)}=0.063841$ and the five-point stress scan gives the coarse crossing estimate $\lambda_{\mathrm{cross}}\approx0.073406$, with the sampled optimum changing from $j=2$ to $j=1$. For C4 $(\theta=0.125,q=x)$, the corresponding values are $\lambda_{\mathrm{step}}^{(1)}=0.015935$ and a coarse crossing estimate $\lambda_{\mathrm{cross}}\approx0.016445$, with $j=6\to5$. By contrast, C1 $(\theta=0.35,q=z)$ and C3 $(\theta=0.15,q=z)$ remain on the preservation side through $1.25\lambda_{\mathrm{step}}^{(1)}$. Three of the 20 HEOM points are therefore reversed, and two of the four candidate families exhibit an observed crossing. Using the sampled-argmax predictor of Eq.~\eqref{eq:first-order-predicted-sampled-argmax} with the same smaller-$j$ tie rule as the HEOM observation, the fixed ideal rule captures $0/3$ observed changes but remains correct on all $17/17$ unchanged points. The first-order predictor captures all $3/3$ observed changes and has no false negatives, but it is correct on only $8/17$ unchanged points because it makes nine early transition calls. This is the intended threshold-stress regime: the candidate families were selected for small first-order thresholds and three of the five ratios lie at or above $\lambda_{\mathrm{step}}^{(1)}$. The HEOM result therefore establishes high sensitivity to actual stopping changes, not uniform all-point superiority; higher-order dynamics is needed to reject early calls near stabilized boundaries.

This experiment changes the role of the threshold theory relative to the broad scan. The production domains establish robust preservation away from a boundary; the targeted scan then demonstrates that genuine stopping transitions emerge as the coupling approaches the theory-selected scale. The first-order threshold is close to quantitative for C4, while C2 and the non-crossing controls expose finite-coupling corrections that are analyzed next.

\subsection{Hierarchy refinement and higher-order stabilization}
\label{subsec:threshold-refinement-results}

The observed crossing locations are next refined to separate physical first-order error from HEOM hierarchy truncation. For C2 and C4, the original sign-change brackets are reevaluated at $(L,N_k)=(3,6),(4,6),(3,8),(4,8)$ and then refined by bisection. All eight hierarchy/candidate combinations retain the original crossing topology. The hierarchy-refined roots are
\[
\lambda_{\mathrm{cross}}^{\mathrm{C2}}=0.072700,
\qquad
\lambda_{\mathrm{cross}}^{\mathrm{C4}}=0.016426.
\]
Relative to first order, these correspond to shifts of $+13.8762\%$ for C2 and $+3.0816\%$ for C4. The cross-configuration root spans are only $6.18\times10^{-4}$ and $1.8\times10^{-5}$, respectively, so the C2 delay is far larger than the numerical hierarchy uncertainty, while C4 remains a near-quantitative threshold prediction.

\begin{figure}[t]
    \centering
    \includegraphics[width=0.98\linewidth]{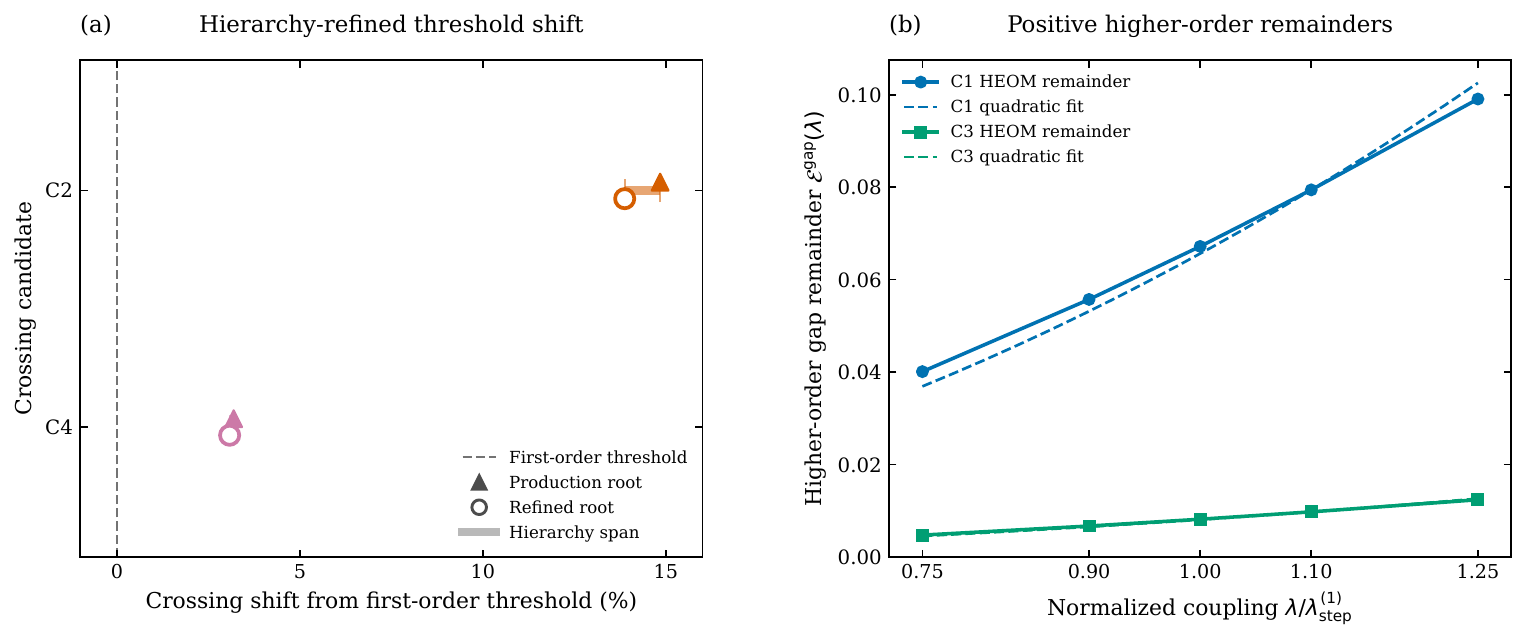}
    \caption{Threshold refinement and higher-order effects. Panel (a) compares the first-order thresholds with production and hierarchy-refined crossing locations for C2 and C4; the hierarchy spans are much smaller than the physical first-order shifts. Panel (b) shows the positive higher-order remainders for the non-crossing controls C1 and C3 together with local through-origin quadratic diagnostics.}
    \Description{Two-panel plot showing hierarchy-refined crossing shifts for C2 and C4 and positive higher-order remainder curves for C1 and C3.}
    \label{fig:threshold-refinement}
\end{figure}

The non-crossing controls provide the complementary higher-order diagnostic. Using the notation of Appendix~\ref{app:higher-order-threshold}, define
\[
\mathcal E_q^{\mathrm{gap}}(\lambda)
=
\Delta_q^{\mathrm{HEOM}}(\lambda)
-
\bigl[\Delta_0-\lambda D_q\bigr].
\]
For both C1 and C3, $\mathcal E_q^{\mathrm{gap}}(\lambda)$ is positive at all five sampled points. A through-origin quadratic diagnostic $\mathcal E_q^{\mathrm{gap}}(\lambda)\approx a_{2,q}\lambda^2$ has coefficient of determination $0.984561$ for C1 and $0.996039$ for C3. At the pre-specified endpoint $1.25\lambda_{\mathrm{step}}^{(1)}$, all four hierarchy settings also keep the full-HEOM sampled gap positive. Figure~\ref{fig:threshold-refinement} therefore separates numerical root refinement from genuine finite-coupling stabilization. The quadratic fits are local diagnostics, not an exact global second-order law.

\subsection{Bath robustness under recalibration and fixed operating coupling}
\label{subsec:physical-robustness-results}

We use two complementary bath-robustness tests because they answer different operational questions. The first retains the C1--C4 threshold families but recalibrates $\lambda_{\mathrm{step}}^{(1)}$ separately at five bath settings: baseline $(\gamma,T)=(0.5,1.0)$, $\gamma\pm20\%$, and $T\pm20\%$. Full HEOM is then evaluated at the same five normalized ratios $\lambda/\lambda_{\mathrm{step}}^{(1)}\in\{0.75,0.90,1.00,1.10,1.25\}$, giving 100 points. This scan tests whether the predicted crossing topology survives when the operating coordinate is recalibrated with the bath. The qualitative partition is unchanged: C2 and C4 cross in all $5/5$ settings, while C1 and C3 remain non-crossing through the pre-specified upper endpoint in all $5/5$ settings. Across the 100 points, 15 genuine sampled-step reversals occur; the first-order rule detects all $15/15$ but makes 45 conservative early calls on the 85 unchanged points. A separate crossing analysis shows that all ten observed C2/C4 crossings over the five bath settings occur \emph{after} the corresponding first-order threshold, with $\lambda_{\mathrm{cross}}/\lambda_{\mathrm{step}}^{(1)}\in[1.019762,1.166880]$, while C1/C3 remain non-crossing through $1.25\lambda_{\mathrm{step}}^{(1)}$. Thus the zero-false-negative pattern on this grid reflects stabilizing finite-coupling corrections in the tested families, under bath parameters known for threshold recalibration, rather than a general one-sided-error property of the first-order rule. The magnitude of the bath recalibration is operationally non-negligible: for C2, lowering the temperature to $T=0.8$ shifts the first-order threshold from the baseline value $0.063841$ to $0.0783106488$, an increase of approximately $22.7\%$. The targeted C2, $T-20\%$ hierarchy check also preserves the $1.10$--$1.25$ sign bracket when the depth is increased from $L=3$ to $L=4$, so the perturbed crossing topology is not a production-depth artifact.

The second test keeps the \emph{physical operating coupling} fixed while the realized bath changes. For each of C1--C4, the operating points are fixed at $0.90$, $1.00$, and $1.10$ times the \emph{baseline-bath} first-order threshold and are then reevaluated under the same five bath settings, producing 60 physical points. A bath-induced truth change is defined by comparing the HEOM stopping winner at the perturbed bath with the baseline-bath winner at the same candidate and same fixed physical $\lambda$. Among the 48 non-baseline perturbations, six true winners change: one under $\gamma-20\%$, none under $\gamma+20\%$, one under $T-20\%$, and four under $T+20\%$. By family, C1/C3 have no bath-induced winner changes, C2 has two, and C4 has four. Fixed-$\lambda$ bath mismatch is therefore a genuine physical effect rather than only a change of threshold coordinates.

We compare two low-order threshold decisions against the same expanded-$\mathcal J_3$ HEOM truth. The fixed-baseline rule retains the baseline threshold decision, whereas the bath-recalibrated rule keeps the physical $\lambda$ unchanged but recomputes $\lambda_{\mathrm{step}}^{(1)}$ for the realized bath. Over all 60 points, exact-step recovery improves from $27/60=45.0\%$ to $38/60=63.33\%$, and mean stopping regret decreases from $0.0149470$ to $0.00917508$. On the same 60-point HEOM domain, the Fixed-Ideal rule is exact on $51/60=85.0\%$ and has the lowest overall mean stopping regret, $8.23829\times10^{-4}$. The aggregate comparison is dominated by the 54 points for which the bath does not change the HEOM-optimal stopping step; on that subset Fixed Ideal is exact on $50/54$ with mean regret $2.85276\times10^{-4}$. The role of bath recalibration is instead concentrated on the six bath-induced truth-change points: Fixed Ideal is exact on $1/6$, the fixed-baseline rule on $4/6$, and the bath-recalibrated rule on $6/6$ with zero stopping regret. Recalibration changes 17 decisions relative to the fixed-baseline rule: 14 replace an incorrect decision by the correct one, while three replace a previously correct decision by an incorrect one, all under $T+20\%$. Because these three cases lie in the most bath-sensitive direction, we additionally recompute the 12 baseline and 12 $T+20\%$ operating points at hierarchy depth $L=4$ with fixed $N_k=6$. All $24/24$ stopping labels and signed-gap signs agree with their $L=3$ values; the same four $T+20\%$ bath-induced truth changes are preserved, and the same three cases still change from a correct fixed-baseline decision to an incorrect recalibrated decision. This is a targeted hierarchy-depth robustness check, not a claim of full HEOM hierarchy convergence. The fixed-operating-$\lambda$ test therefore supports a regime-dependent interpretation rather than an overall superiority claim.

\begin{figure}[t]
    \centering
    \includegraphics[width=0.98\linewidth]{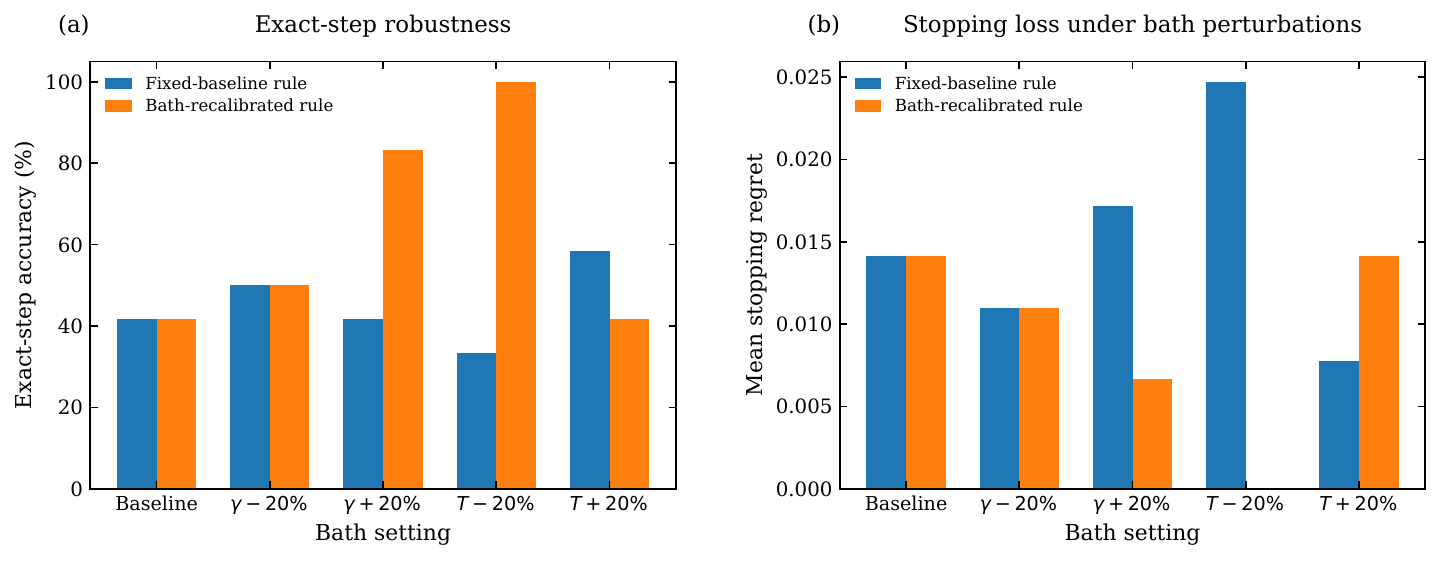}
    \caption{Fixed-operating-$\lambda$ bath robustness on the threshold-centered 60-point stress set. Panel (a) compares exact-step recovery for the fixed-baseline rule and the bath-recalibrated rule under the baseline bath and independent $\pm20\%$ variations of $\gamma$ and $T$. Panel (b) reports the corresponding mean stopping regret. The physical operating coupling is held fixed under both rules: the fixed-baseline rule retains the baseline-bath threshold decision, whereas the bath-recalibrated rule recomputes the first-order threshold for the realized bath. Relative to the fixed-baseline threshold decision, recalibration improves aggregate recovery and regret, but the $T+20\%$ cases show that the correction is not pointwise monotone.}
    \Description{Two-panel fixed-operating-coupling bath-mismatch stress test comparing exact-step accuracy and mean stopping regret before and after bath-threshold recalibration.}
    \label{fig:fixed-lambda-bath-robustness}
\end{figure}

Figure~\ref{fig:fixed-lambda-bath-robustness} summarizes the fixed-operating-$\lambda$ stress test. It is intentionally threshold-centered rather than a broad-domain accuracy benchmark: the purpose is to expose operating-point sensitivity near the transition scale. Together, the two tests distinguish \emph{recalibrated threshold-topology robustness} from \emph{fixed-operating-point mismatch robustness}. The former shows that the crossing/stabilization topology survives the tested $\pm20\%$ bath perturbations after re-expression in the locally recalibrated threshold coordinate; it should not be read as a claim of global robustness at an unchanged physical coupling. The latter directly tests that fixed-coupling question and shows that bath-aware recalibration improves the average decision while offering no pointwise guarantee.

\subsection{Aggregate performance and screening--escalation workflow}
\label{subsec:aggregate-screening-results}

The five HEOM validation datasets contain 346 dataset instances, but the coupling hold-out, threshold, bath, and broad scans overlap in physical parameter space. After cross-dataset deduplication they correspond to 260 unique physical HEOM points, of which 15 exhibit a genuine stopping-step change relative to the ideal sampled optimum and 245 remain unchanged. The 15 changing points arise from the two crossing families C2/C4 across the five tested bath settings, rather than from 15 independent family-level events. We therefore use the unique-point domain for aggregate classification and regret, while retaining dataset-instance rows only for dataset-specific operational accounting. A positive case means $j_q^*\neq j_0^*$, and the sampled-success stopping regret of a rule selecting $j_{\mathrm{rule}}$ is
\[
    R_{\mathrm{stop}}
    =
    P_{\mathrm{HEOM}}(j_q^*\tau_G)
    -
    P_{\mathrm{HEOM}}(j_{\mathrm{rule}}\tau_G).
\]

Table~\ref{tab:aggregate-screening-performance} shows the class-imbalance effect explicitly on the deduplicated physical-point domain: the fixed ideal rule attains high raw accuracy by preserving the dominant unchanged class but misses every genuine change, whereas the first-order rule detects all 15 observed changes with no false negatives at the cost of 45 conservative false positives. The first-order layer is therefore evaluated as a high-sensitivity screen rather than by pooled accuracy or regret alone.

\begin{table}[t]
\caption{Aggregate stopping performance over 260 unique physical HEOM points. Precision/recall treat a genuine HEOM stopping-step change relative to $j_0^*$ as the positive class.}
\label{tab:aggregate-screening-performance}
\small
\centering
\begin{tabular}{@{}lcccccc@{}}
\toprule
Method & Exact step & Precision & Recall & Specificity & Balanced acc. & Mean regret \\
\midrule
Fixed ideal & $245/260$ & -- & $0.000$ & $1.000$ & $0.5000$ & $4.5234\times10^{-4}$ \\
First order & $215/260$ & $0.250$ & $1.000$ & $0.8163$ & $0.9082$ & $4.1634\times10^{-3}$ \\
\bottomrule
\end{tabular}
\end{table}

The regret decomposition makes the complementary failure modes explicit. On the 15 unique genuine-change points, mean stopping regret is $7.84063\times10^{-3}$ for Fixed Ideal and zero for First order. On the 245 stable points, the corresponding means are zero and $4.41833\times10^{-3}$. All 45 non-exact first-order decisions are exactly one sampled step early and none are late; their mean regret is $2.40554\times10^{-2}$. Detailed subgroup statistics are recorded in Appendix~\ref{app:screening-workflow-protocol}.

This error structure motivates an explicit screening--escalation workflow. We use the pre-specified threshold $\eta_{\mathrm{esc}}=0.75=1-\delta_{\mathrm{esc}}$ with safety buffer $\delta_{\mathrm{esc}}=0.25$ below the nominal first-order transition at $\eta_{\mathrm{step}}^{(1)}=1$. This buffer is an operational design choice motivated by the separated-margin form of Proposition~\ref{prop:first-order-sampled-cell}, not a theorem-level certificate and not a value tuned after observing the validation outcomes. A point is accepted at low order only when the conservative first-order sampled-gap indicator lies below $\eta_{\mathrm{esc}}$ and the first-order rule preserves $j_0^*$; otherwise the point is escalated to finite-$\lambda$ HEOM, which acts as the final arbiter. The values $\eta_{\mathrm{esc}}=0.70$ and $0.80$ are used only as sensitivity checks and are not re-optimized after observing the outcomes.

At the primary threshold $\eta_{\mathrm{esc}}=0.75$, the deduplicated workflow accepts 160 of the 260 unique physical points at low order and escalates 100. All 15 unique genuine stopping changes are escalated, no changed point is incorrectly accepted as stable, and resolving the escalated points with full HEOM returns the exact HEOM sampled optimum on all 260 unique points. Thus a full-HEOM-everywhere strategy requires 260 unique physical propagations whereas the screening workflow requires 100, avoiding 160 propagations ($61.54\%$) within this pre-specified validation set. The corresponding dataset-instance accounting is 226 low-order acceptances and 120 escalations over 346 overlapping rows; those row counts are retained only as operational provenance, not as independent physical validations. The low-order screening basis itself requires 44 unique linearized-response solves.

A same-environment timing benchmark on five representative validated configurations measures median runtimes of $0.0199159$~s for one production full-HEOM stopping propagation and $0.3433415$~s for one production linearized-response solve, giving a median runtime ratio $r=17.2396$. Using this measured ratio only as a coarse implementation-level proxy gives
\[
    \frac{C_{\mathrm{workflow}}}{C_{\mathrm{full}}}
    \approx
    \frac{100+44r}{260}
    =3.3021.
\]
Three reasonable timing aggregations give ratios from $3.2343$ to $3.3021$. Under the present implementation and benchmark conditions, the estimated workflow/full wall-clock cost ratio is therefore approximately $3.30$, so the reduction in full-HEOM evaluations does not yield a net wall-clock saving. A likely contributor to this implementation-level overhead is that the current linearized routine constructs a 2401-point response trajectory, whereas the full-HEOM stopping evaluation propagates only at the sparse sampled times required by $\mathcal J_3$. This timing comparison is an implementation-level cost estimate rather than an asymptotic complexity claim.

\begin{table}[t]
\caption{Dataset-level stopping and primary screening performance at $\eta_{\mathrm{esc}}=0.75$. Exact-step columns are scored against HEOM; ``False stable'' counts genuine HEOM changes incorrectly accepted without escalation. Dataset rows overlap in physical parameter space and are retained for provenance rather than aggregate physical-point statistics.}
\label{tab:screening-escalation}
\scriptsize
\centering
\begin{tabular}{@{}lrrrrrr@{}}
\toprule
Dataset & $N$ & HEOM changes & Fixed ideal & First order & Escalate & False stable \\
\midrule
Production & 108 & 0 & $108/108$ & $108/108$ & 0 & 0 \\
Coupling hold-out & 54 & 0 & $54/54$ & $54/54$ & 0 & 0 \\
Small angle & 64 & 0 & $64/64$ & $64/64$ & 0 & 0 \\
Threshold stress & 20 & 3 & $17/20$ & $11/20$ & 20 & 0 \\
Bath-recalibrated stress & 100 & 15 & $85/100$ & $55/100$ & 100 & 0 \\
\bottomrule
\end{tabular}
\end{table}

All broad-set screening values are at most $0.6410034$, whereas the stress grids start at $0.75$, so no tested point lies in $(0.6410034,0.75)$. Because low-order acceptance is strict, points at the $0.75$ boundary are escalated by the primary rule. Raising $\eta_{\mathrm{esc}}$ to $0.80$ accepts 20 additional stable unique physical points while still escalating all $15/15$ unique genuine changes, reducing the unique HEOM workload to $80/260$; the primary pre-specified value remains $0.75$.

\FloatBarrier
\section{Model Discrimination, Circuit-Level Comparison, and Transfer Structure}
\label{sec:comparisons}

\subsection{Same-bath NZ2 threshold discrimination}
\label{subsec:model-comparison-results}

The model-discrimination layer uses the same-bath second-order Nakajima--Zwanzig model (NZ2) defined in Appendix~\ref{app:reference-model-details}, with the same system Hamiltonian, coupling operator, Drude--Lorentz bath, and physical parameters as HEOM.

We reuse the four candidate families C1--C4 and the five fixed ratios $\lambda/\lambda_{\mathrm{step}}^{(1)}\in\{0.75,0.90,1.00,1.10,1.25\}$ from Section~\ref{subsec:threshold-crossing-results}. All 20 NZ2 points pass the numerical reliability checks. Full HEOM crosses only for C2 and C4, whereas NZ2 predicts a crossing in all four families. On the three HEOM change points, both first order and NZ2 are correct at $3/3$. On the 17 unchanged points, NZ2 improves agreement from the first-order $8/17$ to $14/17$, reducing early calls from nine to three. Those three NZ2 false positives have two forms: C1 and C3 cross only at the $1.25\lambda_{\mathrm{step}}^{(1)}$ endpoint where full HEOM remains stabilized, while C2 crosses already at $1.10\lambda_{\mathrm{step}}^{(1)}$ before the full-HEOM transition. NZ2 therefore improves finite-$\lambda$ discrimination, but it serves only as an optional escalation comparator: each operating point requires a separate memory-kernel propagation, and NZ2 does not replace the explicit first-order transition scale or the remainder-controlled criterion of Proposition~\ref{prop:first-order-sampled-cell}. If the first-order and NZ2 decisions disagree, NZ2 is not promoted automatically to final arbiter; the point remains escalated to full HEOM. On the pre-specified 20-point NZ2 set there are six such disagreements, and HEOM supports the NZ2 decision in all six, an empirical result specific to this stress set rather than a general arbitration rule.

Table~\ref{tab:nz2-threshold-performance} summarizes the three rules on this same 20-point HEOM stress domain. Fixed ideal and NZ2 both attain $17/20$ exact-step agreement, but for different reasons: fixed ideal predicts no changes and therefore has zero recall, whereas NZ2 retains recall $1.000$ on the three genuine changes while reducing the first-order false positives from nine to three. The corresponding mean stopping regrets are evaluated against the same HEOM sampled-success truth.

\begin{table}[t]
\caption{Same-domain stopping performance on the 20-point threshold stress set. Precision and recall treat the three genuine HEOM stopping-step changes as positives; regret is scored against the HEOM sampled-success optimum.}
\label{tab:nz2-threshold-performance}
\small
\centering
\begin{tabular}{@{}lcccc@{}}
\toprule
Method & Exact step & Precision & Recall & Mean regret \\
\midrule
Fixed ideal & $17/20$ & -- & $0.000$ & $1.1730\times10^{-3}$ \\
First order & $11/20$ & $0.250$ & $1.000$ & $1.0826\times10^{-2}$ \\
NZ2 & $17/20$ & $0.500$ & $1.000$ & $2.7109\times10^{-3}$ \\
\bottomrule
\end{tabular}
\end{table}

For the genuine crossing families, the same-bath reduction reproduces the transition scale but tends to cross early. The refined NZ2 root for C4 is
\[
\lambda_{\mathrm{cross}}^{\mathrm{NZ2}}=0.016097
\]
versus $\lambda_{\mathrm{cross}}^{\mathrm{HEOM}}=0.016426$, a relative difference of $1.9985\%$. For C2, NZ2 gives $0.065163$ versus the HEOM root $0.072700$, a $10.3678\%$ early transition. The two non-crossing controls expose the complementary failure mode: at the pre-specified endpoint $1.25\lambda_{\mathrm{step}}^{(1)}$, C1 has normalized sampled gaps $+0.209043$ (HEOM) and $-0.004631$ (NZ2), while C3 has $+0.047305$ and $-0.104176$, respectively. Thus, on this pre-specified 20-point stress set, NZ2 captures the genuine reversals in C2/C4 but does not retain the additional stabilization seen in full HEOM for C1/C3. This is a model-discrimination result within the tested bath and candidate families, not a general ordering of NZ2 and HEOM accuracy.

\begin{figure}[t]
    \centering
    \includegraphics[width=0.98\linewidth]{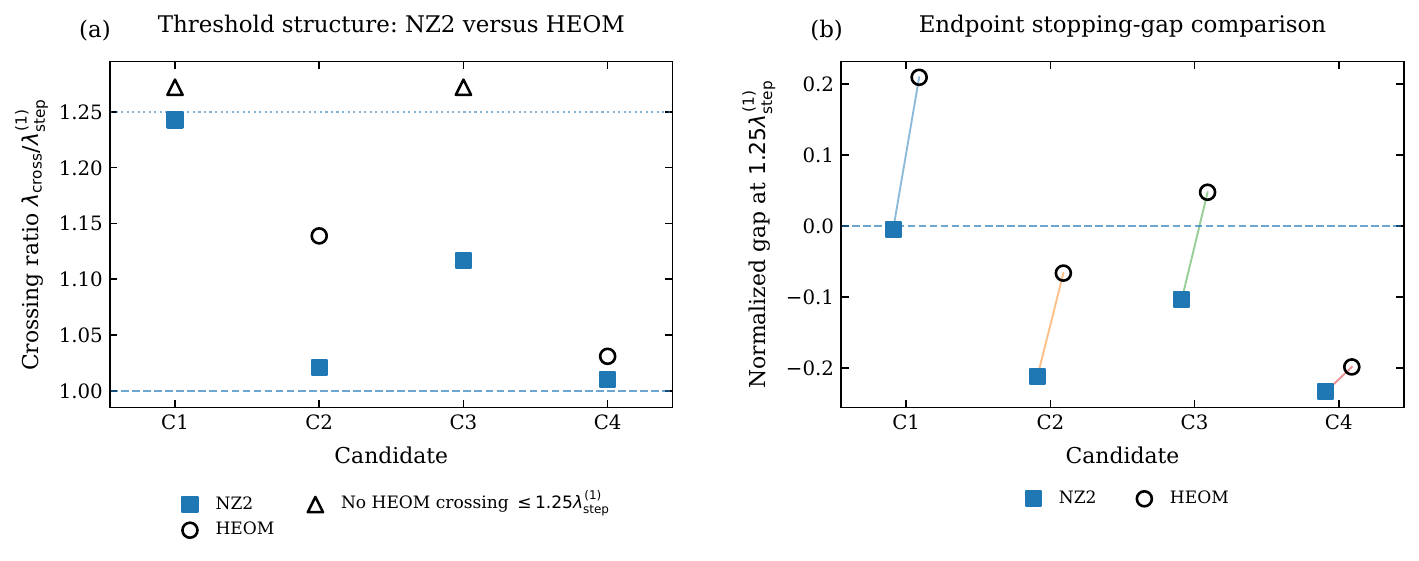}
    \caption{Same-bath NZ2--HEOM threshold comparison. Panel (a) compares candidate-level crossing structure under the common sampled stopping rule: C2 and C4 cross in both models, while C1 and C3 cross only in NZ2 over the tested range. Panel (b) compares the endpoint sampled gaps for all four candidates; C1 and C3 are the HEOM-stabilized controls and expose the NZ2 false-positive crossings.}
    \Description{Two-panel comparison of same-bath NZ2 and full-HEOM stopping-threshold behavior for the four candidate families, including the non-crossing C1 and C3 controls.}
    \label{fig:nz2-heom}
\end{figure}

The continuous NZ2--HEOM timing residual also remains separated from the propagated numerical floor throughout the broad primary regime, even where the sampled stopping step is unchanged. Figure~\ref{fig:nz2-heom} summarizes the threshold-level comparison.

\subsection{Circuit-level stopping comparison}
\label{subsec:circuit-validation}

A separate $n=3$ Grover circuit test probes the sampled-response layer under heterogeneous device-inspired noise. All eight marked targets are compiled once to a seeded three-qubit Qiskit fake backend with restricted connectivity, qubit-dependent coherence times, heterogeneous gate errors and durations, and readout error. One integer circuit iteration is an oracle followed by diffusion, and the circuit stopping optimum is taken over $\mathcal J_{\mathrm{circ}}=\{0,1,2,3\}$. The exact density-matrix scan excludes readout error to isolate sampled ordering, whereas raw readout error is included only in the finite-shot layer. The dimensionless scale $s$ multiplies the fixed seeded device-noise calibration, with $s=1$ the nominal fake-device calibration and $s>1$ used only as a controlled stress test. The circuit branch constructs its own local first-order response from exact density-matrix simulations at small noise, independently of the HEOM reduced model; no finite-$s$ optimum is fitted.

Across the 56 exact target--scale points, the first-order rule agrees with the exact noisy sampled optimum at $48/56=85.71\%$, compared with $32/56=57.14\%$ for the fixed ideal rule. The exact stopping transition occurs at the three scales $s=2.0,2.5,3.0$, with the same $j=2\to1$ change replicated across all eight marked targets at each scale (24 target--scale instances). The first-order rule recovers all three transition scales across all eight targets, whereas the fixed ideal rule recovers none of these transition instances. Conversely, the eight first-order false positives are the single intermediate scale $s=1.5$ replicated across all targets, where the rule calls the $j=2\to1$ transition one scale before the exact sampled optimum. The sampled-success cost of this trade-off is small on the exact circuit grid: mean stopping regret is $8.2596\times10^{-4}$ for the first-order rule versus $2.2741\times10^{-2}$ for the fixed ideal rule, so the first-order mean regret is $3.63\%$ of the fixed-ideal value.

The finite-shot layer covers all 56 target--scale groups with 4096 shots and 100 deterministic seeds per group, giving 5600 stopping decisions. Against the exact density-matrix stopping optimum, finite-shot exact-step recovery is $96.18\%$ overall, but falls to $74.5\%$ at the pre-identified $s=1.5$ boundary and is $99.7917\%$ elsewhere. At $s=1.5$, the exact top-two sampled-probability gap has median $0.005780852$ and range $0.004584781$--$0.006980508$ across the eight targets, directly identifying a near-degenerate decision boundary. The reported recovery rate includes both finite-shot sampling fluctuations and unmitigated readout error; their individual contributions are not separated. Detailed gap correlations, error directions, and regret statistics are reported in Appendix~\ref{app:circuit-protocol}.

The response coefficient $\chi_{\mathrm{circ}}$ remains a separate exact-density finite-difference quantity: the two tested perturbation scales agree in sign on all $32/32$ target--depth points, and finite-shot frequencies are never used to define or estimate the response.

\begin{figure}[!htbp]
    \centering
    \includegraphics[width=0.98\linewidth]{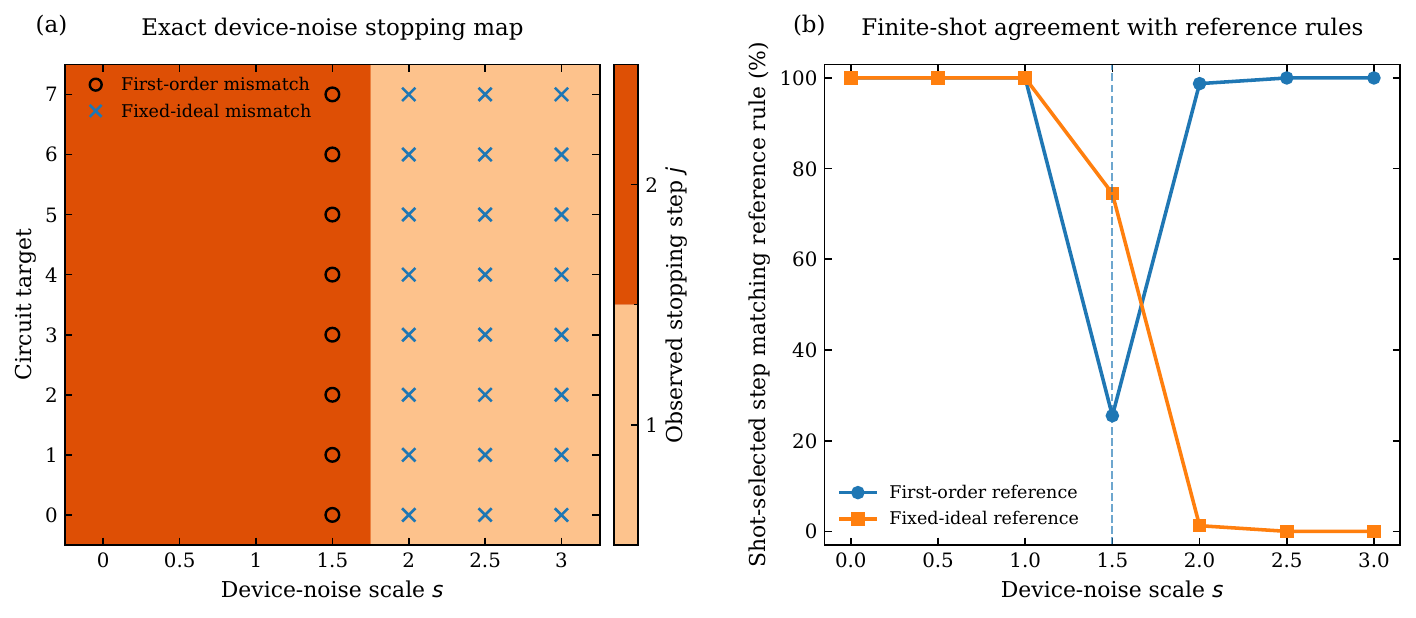}
    \caption{Circuit-level stopping comparison under heterogeneous device-inspired noise. Panel (a) shows the exact device-noise sampled optimum across eight targets and seven noise scales; open circles and crosses mark first-order and fixed-ideal mismatches, respectively. Panel (b) reports, at each noise scale, the percentage of finite-shot selected steps matching the First-order and Fixed-Ideal reference selections over 5600 deterministic finite-shot decisions. The pronounced $s=1.5$ disagreement boundary is localized rather than representative of the remaining scale range.}
    \Description{Two-panel circuit comparison showing the exact noisy stopping map with rule mismatches and, by noise scale, the percentage of finite-shot selected steps matching the First-order and Fixed-Ideal reference selections.}
    \label{fig:circuit-comparison}
\end{figure}

Figure~\ref{fig:circuit-comparison} separates the exact stopping map from the finite-shot stability layer. The open circles and crosses in panel (a) mark disagreements of the first-order and fixed-ideal rules, respectively, with the exact noisy optimum. Panel (b) instead reports, at each noise scale, the percentage of finite-shot selected steps that match the First-order and Fixed-Ideal reference selections, making the $s=1.5$ disagreement boundary explicit.

The Leng comparison is evaluated on the same 56 exact-density trajectories under both objectives. For sampled success, the first-order rule agrees with the exact argmax on $48/56$ points with mean success regret $8.2596\times10^{-4}$, whereas the Leng-selected depth agrees on $24/56$ points with mean success regret $4.54191\times10^{-2}$. Under the exact restart-aware runtime objective, Leng has zero mean runtime regret, whereas the first-order rule has mean runtime regret $0.448623$. The methods select the same depth on 32 points and differ on 24. On every disagreement point, the first-order selection has lower sampled-success regret and the Leng selection has lower restart-aware runtime regret. This difference reflects the two objectives on this particular grid rather than evidence of broad method superiority: the exact runtime optimum is $j=1$ at all 56 points, and the Leng rule also selects $j=1$ throughout. Fidelity values and the objective definitions are given in Appendix~\ref{app:small-angle-protocol}.

\subsection{Local quadratic transfer-loss relation}
\label{subsec:transfer-loss-results}

The transfer analysis returns to the local continuous first-peak structure of Proposition~\ref{prop:quadratic-transfer-loss}. At $\theta=0.50$, the full eight-point grid $\lambda=0.01,\ldots,0.08$ is evaluated for both coupling geometries and both transfer directions. We use the mean cross-response loss
\[
R_{\mathrm{cross}}
=
\frac{1}{2}\left(R_{z\leftarrow x}+R_{x\leftarrow z}\right),
\qquad
\Delta t_{xz}^{*}=t_x^{*}-t_z^{*},
\]
and regress it through the origin against the normalized timing coordinate
\[
X
=
\left(\frac{|\Delta t_{xz}^{*}|}{\tau_G}\right)^2,
\qquad
\tau_G=\frac{4\theta}{\Omega_G}.
\]
For the numerical convention $\Omega_G=1$ and $\theta=0.50$, one has $\tau_G=2$ and ideal curvature $\kappa_0=1/2$, so Proposition~\ref{prop:quadratic-transfer-loss} gives the non-fitted local coefficient
\[
s_{\mathrm{th}}
=
\frac{1}{2}\kappa_0\tau_G^2
=
1.
\]

The through-origin fits give
\[
s_{\mathrm{fit}}(L=3)=0.996106003,
\qquad
s_{\mathrm{fit}}(L=6)=0.996296647,
\]
with coefficients of determination $0.999623466$ and $0.999548831$, respectively. The depth-refined $L=6$ coefficient differs from the theoretical value by only $0.3703\%$, and the fitted slope changes by only $0.0191\%$ between $L=3$ and $L=6$. At the strongest coupling $\lambda=0.08$, the channel timing mismatch and transfer loss remain more pointwise depth-sensitive, changing by $8.844\%$ and $17.036\%$; the latter is consistent with the square of the former, so these pointwise changes move along essentially the same local quadratic relation rather than changing its fitted coefficient. The validation is intentionally local to the tested first-peak neighborhood and coupling range; it is not evidence for a global loss law far from the peak or across arbitrary noise regimes.

\begin{figure}[!htbp]
    \centering
    \includegraphics[width=0.98\linewidth]{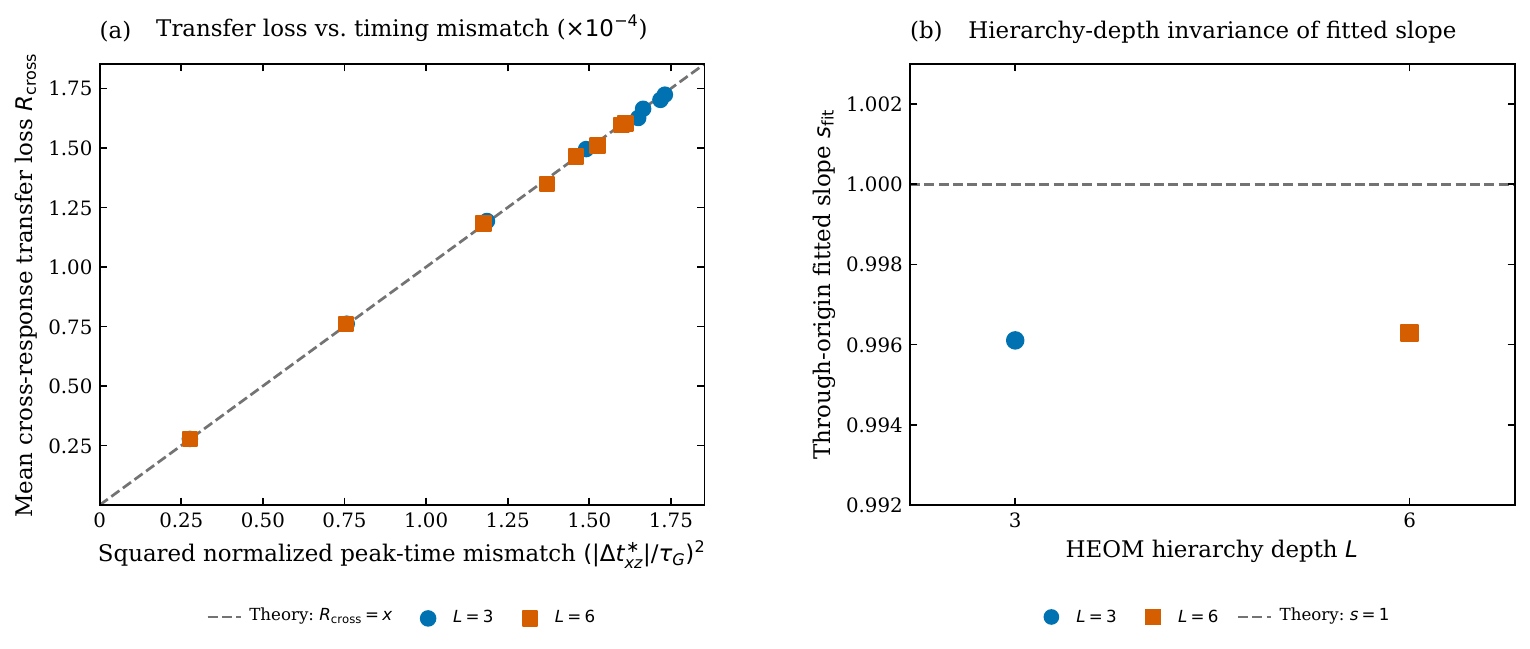}
    \caption{Local quadratic transfer-loss relation. Panel (a) plots the mean cross-response transfer loss against the squared normalized peak-time mismatch in the tested first-peak regime; both axes are displayed in units of $10^{-4}$. Panel (b) shows the through-origin fitted slopes at hierarchy depths $L=3$ and $L=6$ against the non-fitted local theoretical coefficient $s_{\mathrm{th}}=1$.}
    \Description{Two-panel transfer-loss validation showing an approximately unit-slope local quadratic timing-to-loss relation and its stability under HEOM hierarchy-depth refinement.}
    \label{fig:transfer-loss}
\end{figure}

Figure~\ref{fig:transfer-loss} shows this local structural relation. For compactness, both axes in panel (a) are displayed in units of $10^{-4}$.

\FloatBarrier

\section{Practical Implications and Scope}
\label{sec:implications}

The framework functions as a two-stage stopping workflow: well-separated cases are screened at low order, while points approaching the sampled-ordering transition are escalated to finite-$\lambda$ dynamics. The first-order threshold therefore serves as a vulnerability scale for the discrete stopping decision, not as a pointwise surrogate for the full HEOM trajectory. The observed post-threshold C2/C4 crossings and stabilized C1/C3 controls show why this separation is useful: low-order information identifies where omitted finite-coupling structure becomes decision-relevant. Computationally, the workflow reduces the number of full-HEOM stopping evaluations, but the same-environment benchmark in Section~\ref{subsec:aggregate-screening-results} shows that the current dense linearized-response implementation does not reduce total wall-clock cost. The benchmark characterizes the present direct implementation rather than an optimized screening code path; more task-specific sparse or adaptive response evaluation could reduce this overhead, but engineering such an optimized implementation is beyond the scope of this study.

The two bath studies isolate complementary operational questions. Recalibrating the threshold with the realized bath tests whether the crossing topology survives in locally rescaled coordinates, whereas holding the physical coupling fixed tests parameter mismatch at an unchanged operating point. The latter shows that bath-aware recalibration is most useful when the bath actually changes the HEOM-optimal stopping decision, while the full-grid aggregate still favors Fixed Ideal. The targeted hierarchy-depth check preserves the relevant $T+20\%$ decision labels, supporting this regime-dependent interpretation.

The comparison layers clarify how the workflow can be extended without changing its basic logic. Same-bath NZ2 improves finite-$\lambda$ discrimination on the threshold stress set but remains an optional intermediate comparator before HEOM arbitration. The independent circuit branch shows that the sampled-response idea transfers beyond the reduced HEOM model: the exact transition appears at three noise scales across all eight targets, and finite-shot failures concentrate near the narrow-gap $s=1.5$ boundary. The Leng comparison is retained as an objective-specific consistency test on the same circuit trajectories, while the depth-refined transfer experiment supports the predicted local quadratic timing-to-loss relation. The present numerical scope is the tested Grover-active Drude--Lorentz model and a small seeded three-qubit circuit model; broader environments, larger amplitude-amplification settings, and hardware studies remain natural extensions.

\section{Conclusion}
\label{sec:conclusion}

We developed a response-threshold framework for the integer stopping layer of noisy Grover search under the standard calibration $\Omega_G\tau_G=4\theta$. An a priori linearized response provides the first-peak timing coefficient, sampled-gap differentials, and a first-order transition scale, while finite-$\lambda$ dynamics are reserved for parameter points whose sampled ordering approaches that scale. The three-point candidate rule is independently validated against the complete first Grover lobe on the tested HEOM domain.

The numerical evidence supports the screening interpretation. Broad domains remain on the preservation side, while the threshold-focused sets contain both genuine stopping changes and higher-order-stabilized controls. The primary screen escalates every observed unique stopping change in the tested set. Bath variation further distinguishes recalibrated threshold topology from fixed-operating-point mismatch, and the targeted depth check preserves the qualitative decisions in the most bath-sensitive tested direction.

The same-bath NZ2 comparison, the independent circuit experiment, and the depth-refined transfer test delineate the framework's transferability and limits. The circuit branch recovers the observed transition scales and localizes finite-shot instability to a narrow decision boundary, while the Leng comparison separates sampled-success and restart-aware runtime objectives on the same exact-density trajectories. Together, these results position the framework as a stopping-layer screening theory whose next extensions are an explicit higher-order screening response, broader structured environments, larger amplitude-amplification instances, and hardware models with leakage and device-level effects.

\section*{Data and Code Availability}

The data and code supporting the numerical results in this paper are supplied with this submission in the accompanying reproducibility artifact. The package includes the HEOM and linearized-HEOM scripts, sampled-threshold analyses, parameter settings, processed figure data, deduplicated unique-point statistics, convergence and bath-robustness checks, targeted hierarchy-depth checks, numerical-protocol and compute-cost benchmark records, NZ2 comparisons, device-inspired circuit simulations, pointwise incremental-channel fidelities and First-order/Leng dual-objective comparison tables, and plotting code. No human-subject, confidential, or restricted data are used in this study.

\section*{Declaration of Generative AI and AI-Assisted Technologies}

The authors used OpenAI ChatGPT during manuscript preparation to assist with language polishing, manuscript organization, LaTeX formatting, and consistency checking. The tool was not used to generate the numerical data or figures. All AI-assisted text was reviewed, edited, and verified by the authors. The authors take full responsibility for the accuracy, integrity, and originality of the manuscript.

\begin{acks}
The authors thank the Faculty of Information Technology at Monash University for research support.
\end{acks}

\bibliographystyle{ACM-Reference-Format}
\bibliography{references}

@inproceedings{grover1996fast,
  author    = {Lov K. Grover},
  title     = {A Fast Quantum Mechanical Algorithm for Database Search},
  booktitle = {Proceedings of the Twenty-Eighth Annual ACM Symposium on Theory of Computing},
  pages     = {212--219},
  year      = {1996},
  publisher = {ACM},
  doi       = {10.1145/237814.237866}
}

@article{boyer1998tight,
  author  = {Michel Boyer and Gilles Brassard and Peter H{\o}yer and Alain Tapp},
  title   = {Tight Bounds on Quantum Searching},
  journal = {Fortschritte der Physik},
  volume  = {46},
  number  = {4--5},
  pages   = {493--505},
  year    = {1998},
  doi     = {10.1002/(SICI)1521-3978(199806)46:4/5<493::AID-PROP493>3.0.CO;2-P}
}

@incollection{brassard2002quantum,
  author    = {Brassard, Gilles and H{\o}yer, Peter and Mosca, Michele and Tapp, Alain},
  title     = {Quantum amplitude amplification and estimation},
  booktitle = {Quantum Computation and Information},
  series    = {Contemporary Mathematics},
  volume    = {305},
  pages     = {53--74},
  publisher = {American Mathematical Society},
  year      = {2002},
  doi       = {10.1090/conm/305/05215}
}

@book{nielsen2010quantum,
  author    = {Michael A. Nielsen and Isaac L. Chuang},
  title     = {Quantum Computation and Quantum Information},
  edition   = {10th Anniversary},
  publisher = {Cambridge University Press},
  address   = {Cambridge, U.K.},
  year      = {2010}
}

@article{tanimura1989time,
  author  = {Yoshitaka Tanimura and Ryogo Kubo},
  title   = {Time Evolution of a Quantum System in Contact with a Nearly Gaussian-Markoffian Noise Bath},
  journal = {Journal of the Physical Society of Japan},
  volume  = {58},
  number  = {1},
  pages   = {101--114},
  year    = {1989},
  doi     = {10.1143/JPSJ.58.101}
}

@article{tanimura2020numerically,
  author  = {Yoshitaka Tanimura},
  title   = {Numerically ``Exact'' Approach to Open Quantum Dynamics: The Hierarchical Equations of Motion ({HEOM})},
  journal = {The Journal of Chemical Physics},
  volume  = {153},
  number  = {2},
  pages   = {020901},
  year    = {2020},
  doi     = {10.1063/5.0011599}
}

@book{weiss2012quantum,
  author    = {Ulrich Weiss},
  title     = {Quantum Dissipative Systems},
  edition   = {4th},
  publisher = {World Scientific},
  address   = {Singapore},
  year      = {2012},
  doi       = {10.1142/8334}
}

@article{azuma2002decoherence,
  author  = {Hiroo Azuma},
  title   = {Decoherence in Grover's Quantum Algorithm: Perturbative Approach},
  journal = {Physical Review A},
  volume  = {65},
  number  = {4},
  pages   = {042311},
  year    = {2002},
  doi     = {10.1103/PhysRevA.65.042311}
}

@article{azuma2005higher,
  author  = {Hiroo Azuma},
  title   = {Higher-Order Perturbation Theory for Decoherence in Grover's Algorithm},
  journal = {Physical Review A},
  volume  = {72},
  number  = {4},
  pages   = {042305},
  year    = {2005},
  doi     = {10.1103/PhysRevA.72.042305}
}

@inproceedings{hoyer2003quantum,
  author    = {Peter H{\o}yer and Michele Mosca and Ronald de Wolf},
  title     = {Quantum Search on Bounded-Error Inputs},
  booktitle = {Automata, Languages and Programming},
  series    = {Lecture Notes in Computer Science},
  volume    = {2719},
  pages     = {291--299},
  publisher = {Springer},
  year      = {2003},
  doi       = {10.1007/3-540-45061-0_25}
}

@inproceedings{regev2008impossibility,
  author    = {Oded Regev and Liron Schiff},
  title     = {Impossibility of a Quantum Speed-Up with a Faulty Oracle},
  booktitle = {Automata, Languages and Programming},
  series    = {Lecture Notes in Computer Science},
  volume    = {5125},
  pages     = {773--781},
  publisher = {Springer},
  year      = {2008},
  doi       = {10.1007/978-3-540-70575-8_63}
}

@article{ishizaki2005quantum,
  author  = {Akihito Ishizaki and Yoshitaka Tanimura},
  title   = {Quantum Dynamics of System Strongly Coupled to Low-Temperature Colored Noise Bath: Reduced Hierarchy Equations Approach},
  journal = {Journal of the Physical Society of Japan},
  volume  = {74},
  number  = {12},
  pages   = {3131--3134},
  year    = {2005},
  doi     = {10.1143/JPSJ.74.3131}
}

@article{shi2009efficient,
  author  = {Qiang Shi and Liping Chen and Guangjun Nan and Rui-Xue Xu and YiJing Yan},
  title   = {Efficient Hierarchical Liouville Space Propagator to Quantum Dissipative Dynamics},
  journal = {The Journal of Chemical Physics},
  volume  = {130},
  number  = {8},
  pages   = {084105},
  year    = {2009},
  doi     = {10.1063/1.3077918}
}

@article{pokharel2024better,
  author  = {Pokharel, Bibek and Lidar, Daniel A.},
  title   = {Better-than-classical Grover search via quantum error detection and suppression},
  journal = {npj Quantum Information},
  volume  = {10},
  pages   = {23},
  year    = {2024},
  doi     = {10.1038/s41534-023-00794-6}
}

@article{abughanem2025characterizing,
  author  = {AbuGhanem, Muhammad},
  title   = {Characterizing Grover search algorithm on large-scale superconducting quantum computers},
  journal = {Scientific Reports},
  volume  = {15},
  pages   = {1281},
  year    = {2025},
  doi     = {10.1038/s41598-024-80188-6}
}

@article{srivastava2025switches,
  author  = {Srivastava, Suryansh and Pati, Arun K. and Chakrabarty, Indranil and Bhattacharya, Samyadeb},
  title   = {Using quantum switches to mitigate noise in Grover's search algorithm},
  journal = {Journal of Physics A: Mathematical and Theoretical},
  volume  = {58},
  number  = {10},
  pages   = {105304},
  year    = {2025},
  doi     = {10.1088/1751-8121/ad9efc}
}

@article{leng2025noise,
  author  = {Leng, Jian and Yang, Fan and Wang, Xiang-Bin},
  title   = {Noise-tolerant {Grover}'s algorithm via success-probability prediction},
  journal = {Physical Review Research},
  volume  = {7},
  pages   = {L012017},
  year    = {2025},
  doi     = {10.1103/PhysRevResearch.7.L012017}
}

@article{botsinis2016quantum,
  author  = {Panagiotis Botsinis and Zunaira Babar and Dimitrios Alanis and Daryus Chandra and Hung Viet Nguyen and Soon Xin Ng and Lajos Hanzo},
  title   = {Quantum Error Correction Protects Quantum Search Algorithms against Decoherence},
  journal = {Scientific Reports},
  volume  = {6},
  pages   = {38095},
  year    = {2016},
  doi     = {10.1038/srep38095}
}

@article{terhal2015quantum,
  author  = {Barbara M. Terhal},
  title   = {Quantum Error Correction for Quantum Memories},
  journal = {Reviews of Modern Physics},
  volume  = {87},
  number  = {2},
  pages   = {307--346},
  year    = {2015},
  doi     = {10.1103/RevModPhys.87.307}
}

@article{fowler2012surface,
  author  = {Austin G. Fowler and Matteo Mariantoni and John M. Martinis and Andrew N. Cleland},
  title   = {Surface Codes: Towards Practical Large-Scale Quantum Computation},
  journal = {Physical Review A},
  volume  = {86},
  number  = {3},
  pages   = {032324},
  year    = {2012},
  doi     = {10.1103/PhysRevA.86.032324}
}

@article{cai2023quantum,
  author  = {Zhenyu Cai and Ryan Babbush and Simon C. Benjamin and Suguru Endo and William J. Huggins and Ying Li and Jarrod R. McClean and Thomas E. O'Brien},
  title   = {Quantum Error Mitigation},
  journal = {Reviews of Modern Physics},
  volume  = {95},
  number  = {4},
  pages   = {045005},
  year    = {2023},
  doi     = {10.1103/RevModPhys.95.045005}
}

@article{bharti2022noisy,
  author  = {Kishor Bharti and Alba Cervera-Lierta and Thi Ha Kyaw and Tobias Haug and Sumner Alperin-Lea and Abhinav Anand and Matthias Degroote and Hermanni Heimonen and Jakob S. Kottmann and Tim Menke and Wai-Keong Mok and Sukin Sim and Leong-Chuan Kwek and Al{\'a}n Aspuru-Guzik},
  title   = {Noisy Intermediate-Scale Quantum Algorithms},
  journal = {Reviews of Modern Physics},
  volume  = {94},
  number  = {1},
  pages   = {015004},
  year    = {2022},
  doi     = {10.1103/RevModPhys.94.015004}
}

\clearpage
\appendix

\section{Notation and Detailed Open-System Realization}
\label{app:heom-details}

\subsection{Compact notation table}

\begin{table}[h]
\caption{Principal stopping-layer notation.}
\label{tab:notation-appendix}
\small
\centering
\begin{tabularx}{\textwidth}{@{}l X@{}}
\toprule
Symbol & Meaning \\
\midrule
$t_0^*,t_q^*$ & Ideal and noisy continuous first-peak times. \\
$\mathcal J_{\mathrm{lobe}}$ & Complete nonnegative ideal first-Grover-lobe index set. \\
$\mathcal J_3$ & Three-point local candidate neighbourhood around the ideal sampled winner, clipped to $\mathcal J_{\mathrm{lobe}}$. \\
$j_0^*,j_q^*$ & Ideal first-lobe sampled winner and noisy local sampled maximizer, respectively. \\
$\chi_q^{\mathrm{lin}}$ & A priori first-order success response from linearized HEOM. \\
$G_q^{\mathrm{lin}}$ & Signed a priori first-peak timing coefficient. \\
$\widehat G_q^{\mathrm{traj}}$ & Independent empirical timing slope from finite-$\lambda$ full HEOM. \\
$m_0^\pm,m_{0,q}^{\mathrm{dir}}$ & One-sided and direction-aware stroboscopic margins. \\
$\Delta_{0,j}$ & Ideal sampled probability gap between $j_0^*$ and competitor $j$. \\
$D_{q,j},D_{q,j}^{\mathrm{lin}}$ & Generic sampled response differential and its a priori linearized-HEOM realization. \\
$\mathcal E_q,\mathcal E^{\mathrm{step}}_{q,j},\mathcal E_q^{\mathrm{gap}}$ & Perturbative trajectory, sampled-ordering, and finite-coupling gap remainders, respectively. \\
$\eta_{\mathrm{cell},q}^{(1)}$ & Continuous-cell geometric indicator. \\
$\eta_{\mathrm{step},q}^{(1)}$ & Discrete sampled-ordering indicator. \\
$j_{q,\mathrm{pred}}^{*(1)}$
& First-order predicted sampled argmax at a fixed parameter point. \\

$R_{c\leftarrow d}$ & Success loss from using response $d$'s first-peak time on response $c$. \\
$R_{\mathrm{cross}}$ & Mean of the two cross-channel transfer losses. \\
$R_{\mathrm{stop}}$ & Sampled-success regret of a stopping rule relative to the HEOM optimum. \\
\bottomrule
\end{tabularx}
\end{table}
\FloatBarrier

\subsection{Full HEOM model}

The total Hamiltonian is
\[
    H_{\mathrm{tot}}=H_S+H_B+H_{SB},
    \qquad
    H_{SB}=S_q\otimes B,
\]
with a thermal bosonic bath and Drude--Lorentz spectral density in Eq.~\eqref{eq:drude-lorentz-spectral-density}. We use units $\hbar=k_B=1$ and write $\beta=1/T$. The Matsubara expansion used in the implementation is
\begin{equation}
    C(t)\simeq\sum_{m=0}^{N_k}c_m e^{-\nu_m t},
    \qquad
    \nu_0=\gamma,
    \qquad
    \nu_m=\frac{2\pi m}{\beta}\quad(m\ge1),
    \label{eq:drude-matsubara-frequencies}
\end{equation}
with
\begin{equation}
\begin{aligned}
    c_0&=\lambda\gamma\!\left[\cot\!\left(\frac{\beta\gamma}{2}\right)-i\right],\\
    c_m&=\frac{4\lambda\gamma}{\beta}\,
    \frac{\nu_m}{\nu_m^2-\gamma^2},\qquad m\ge1.
\end{aligned}
\label{eq:drude-matsubara-coefficients}
\end{equation}
Thus $N_k$ is the highest retained positive Matsubara index; in the generic notation below $M=N_k$, so there are $N_k+1$ exponential components including the Drude pole $m=0$. HEOM introduces auxiliary density operators $\rho_{\mathbf n}$ indexed by $\mathbf n=(n_0,\ldots,n_M)$, with unit vectors $\mathbf e_m$ in hierarchy-index space. For the exponential decomposition above, the unscaled hierarchy used here is
\begin{equation}
\begin{aligned}
    \dot\rho_{\mathbf n}
    ={}&-\left(i\mathcal L_S+\sum_{m=0}^{M}n_m\nu_m\right)\rho_{\mathbf n}
    -i\sum_{m=0}^{M}\left[S_q,\rho_{\mathbf n+\mathbf e_m}\right]\\
    &-i\sum_{m=0}^{M}n_m\left(
    c_m S_q\rho_{\mathbf n-\mathbf e_m}
    -c_m^*\rho_{\mathbf n-\mathbf e_m}S_q
    \right),
    \qquad \mathcal L_S(\cdot)=[H_S,\cdot].
    \label{eq:full-heom-hierarchy}
\end{aligned}
\end{equation}
The physical reduced state is $\rho_{\mathbf 0}\equiv\rho_S$. For the factorized initial condition used in the simulations,
$\rho_{\mathbf 0}(0)=\rho_S(0)$ and $\rho_{\mathbf n\neq\mathbf 0}(0)=0$; terms with any negative hierarchy index are understood to vanish. The numerical hierarchy uses a hard triangular truncation $\sum_m n_m\le L$: upward couplings that would generate an ADO with $\sum_m n_m>L$ are set to zero. No separate terminator or tail-closure correction is applied. All stopping observables are extracted only from $\rho_S(t)$~\cite{tanimura1989time,tanimura2020numerically}.

\subsection{Linearized HEOM response}
\label{app:linearized-heom}

For the Drude--Lorentz expansion the coefficients scale linearly with the coupling strength,
\[
    c_m(\lambda,\gamma,T)=\lambda\bar c_m(\gamma,T).
\]
Writing
\[
    \rho_{S,q}(t;\lambda)=\rho^{(0)}(t)+\lambda\rho_{1,q}(t)+O(\lambda^2),
    \qquad
    \rho_{\mathbf e_m}(t;\lambda)=\lambda A_{m,q}(t)+O(\lambda^2),
\]
with higher hierarchy tiers entering only at $O(\lambda^2)$ and above, the first-tier response $A_{m,q}$ obeys
\begin{equation}
    \dot A_{m,q}
    =-(i\mathcal L_S+\nu_m)A_{m,q}
    -i\left(\bar c_m S_q\rho^{(0)}-\bar c_m^*\rho^{(0)}S_q\right),
    \qquad A_{m,q}(0)=0,
    \label{eq:first-tier-response}
\end{equation}
where $\mathcal L_S(\cdot)=[H_S,\cdot]$. The physical first-order response satisfies
\begin{equation}
    \dot\rho_{1,q}
    =-i[H_S,\rho_{1,q}]
    -i\left[S_q,\sum_m A_{m,q}\right],
    \qquad \rho_{1,q}(0)=0.
    \label{eq:physical-first-order-response}
\end{equation}
The a priori success response is
\[
    \chi_q^{\mathrm{lin}}(t)
    =\operatorname{ReTr}[\ketbra{t}{t}\rho_{1,q}(t)],
\]
and its sampled differential is
\[
    D_{q,j}^{\mathrm{lin}}
    =
    \chi_q^{\mathrm{lin}}(j\tau_G)
    -
    \chi_q^{\mathrm{lin}}(j_0^*\tau_G).
\]
Equations~\eqref{eq:first-tier-response}--\eqref{eq:physical-first-order-response} are solved using only the ideal trajectory and the specified bath parameters. This is the route used to obtain $G_q^{\mathrm{lin}}$ and all sampled $D_{q,j}^{\mathrm{lin}}$ in the numerical results.

\section{Same-Bath NZ2 Comparator}
\label{app:reference-model-details}

For a factorized initial condition $\rho_{\mathrm{tot}}(0)=\rho_S(0)\otimes\rho_B^{\mathrm{th}}$ and zero bath mean, define $\widetilde S_q(t)=e^{iH_St}S_qe^{-iH_St}$. The NZ2 interaction-picture equation is
\begin{equation}
\begin{aligned}
    \frac{d}{dt}\widetilde\rho_{\mathrm{NZ2},q}(t)
    =-\int_0^t ds\,\Big(&C(t-s)[\widetilde S_q(t),\widetilde S_q(s)\widetilde\rho_{\mathrm{NZ2},q}(s)]\\
    &+C^*(t-s)[\widetilde\rho_{\mathrm{NZ2},q}(s)\widetilde S_q(s),\widetilde S_q(t)]\Big).
    \label{eq:nz2-memory-kernel}
\end{aligned}
\end{equation}
HEOM and NZ2 use the same finite exponential representation of $C(t)$ and the same $(\lambda,\gamma,T,S_q)$. No HEOM-derived damping, timing, or phase descriptor is used to calibrate NZ2. Because second-order NZ dynamics does not generally guarantee complete positivity, every trajectory is screened for trace, Hermiticity, state positivity, and first-peak identifiability.

\section{Proofs of the Main Stopping Results}
\label{app:proofs}

\subsection{Proof of Proposition~\ref{prop:first-order-first-peak-response}}

Define $F_q(t,\lambda)=\partial_tP_q(t;\lambda,\theta)$. Joint $C^2$ regularity of $P_q$ makes $F_q$ continuously differentiable in $(t,\lambda)$. The assumptions preceding Proposition~\ref{prop:first-order-first-peak-response} give $F_q(t_0^*,0)=0$ and
$\partial_tF_q(t_0^*,0)=P_0''(t_0^*)=-\kappa_0\neq0$. The implicit-function theorem therefore gives a unique stationary branch $t_q^*(\lambda)$ in a sufficiently small neighbourhood of $t_0^*$ for small $\lambda$; the uniform second-derivative remainder bound and continuity preserve negative curvature, so this branch is a local maximum. Write
\[
    t_q^*=t_0^*+\lambda G_q+O(\lambda^2).
\]
Differentiating Eq.~\eqref{eq:success-weak-coupling} and expanding about $t_0^*$, with the derivative remainder uniformly $O(\lambda^2)$, yields
\[
    0=\lambda\left[G_qP_0''(t_0^*)+\dot\chi_q(t_0^*)\right]+O(\lambda^2).
\]
Using $\kappa_0=-P_0''(t_0^*)>0$ gives Eq.~\eqref{eq:Gq-definition} and hence Eq.~\eqref{eq:first-peak-expansion}.

\subsection{Proof of Proposition~\ref{prop:first-order-sampled-cell}}

Evaluate Eq.~\eqref{eq:success-weak-coupling} at $j_0^*\tau_G$ and $j\tau_G$ and subtract:
\[
    P_q(j_0^*\tau_G)-P_q(j\tau_G)
    =\Delta_{0,j}-\lambda D_{q,j}+\mathcal E^{\mathrm{step}}_{q,j}(\lambda),
\]
where the uniform expansion gives $|\mathcal E^{\mathrm{step}}_{q,j}(\lambda)|\le C_q\lambda^2$ on the stated local interval. If $\Delta_{0,j}-\lambda D_{q,j}>C_q\lambda^2$ for every competitor, then every sampled difference remains positive and $j_0^*$ is the unique maximizer. If instead $\lambda D_{q,j}-\Delta_{0,j}>C_q\lambda^2$ for some competitor, then its sampled probability exceeds that at $j_0^*$, so the ideal maximizer is reversed. The $\delta$-separated forms in Proposition~\ref{prop:first-order-sampled-cell} follow by lower-bounding the first-order margin by $\delta\Delta_{0,j}$.

The midpoint rule follows only in the additional special case of a locally symmetric non-degenerate quadratic peak: equality of two adjacent sampled probabilities then occurs when the continuous maximum reaches their midpoint. The general certificate in the main text does not require that symmetry.

\subsection{Proof of Proposition~\ref{prop:quadratic-transfer-loss}}

Taylor expand $P_c$ about its local maximum $t_c^*$:
\[
    P_c(t_c^*+\delta t)
    =P_c(t_c^*)-\frac12\kappa_c\delta t^2+O(|\delta t|^3).
\]
Setting $\delta t=t_d^*-t_c^*$ and using Eq.~\eqref{eq:transfer-loss-definition} yields Eq.~\eqref{eq:quadratic-transfer-loss}.

\subsection{Response-aware calibration bound}
\label{app:calibration-bound}

If an estimated coefficient $\widehat G_q$ satisfies $|\widehat G_q-G_q|\le\epsilon_G$ and the response-aware stopping time is $\widehat t_q=t_0^*+\lambda\widehat G_q$, then
\[
    |\widehat t_q-t_q^*|\le\lambda\epsilon_G+O(\lambda^2).
\]
Combining this with Proposition~\ref{prop:quadratic-transfer-loss} gives
\[
    P_q(t_q^*)-P_q(\widehat t_q)
    \le\frac12\kappa_q\lambda^2\epsilon_G^2+O(\lambda^3),
\]
provided $\widehat t_q$ remains in the local first-peak neighbourhood.


\section{Uncertainty Propagation and Higher-Order Details}
\label{app:uncertainty-details}

Using the uncertainty notation of Section~\ref{subsec:uncertainty-main}, and additionally writing
$|\widehat m_0^\pm-m_0^\pm|\le\epsilon_m^\pm$ for the one-sided stroboscopic margins, the continuous-cell bounds are
\[
    \eta_{\mathrm{cell},q}^{-}
    =
    \frac{\lambda(|\widehat G_q|-\epsilon_G)}
         {\widehat m_{0,q}^{\mathrm{dir}}+\epsilon_{m,q}^{\mathrm{dir}}},
    \qquad
    \eta_{\mathrm{cell},q}^{+}
    =
    \frac{\lambda(|\widehat G_q|+\epsilon_G)}
         {\widehat m_{0,q}^{\mathrm{dir}}-\epsilon_{m,q}^{\mathrm{dir}}}.
\]
If the response sign is unresolved, the conservative upper bound uses the smaller admissible one-sided margin and the lower bound is set to zero. Equation~\eqref{eq:uncertain-sampled-step-indicator-bounds} follows by combining the most destabilizing admissible numerator with the smallest admissible positive denominator, and conversely for the lower bound.

The numerical budgets are deterministic rather than repeated-sampling errors. For $G_q^{\mathrm{lin}}$ we use
\[
    \epsilon_G
    =
    \epsilon_G^{\mathrm{trunc}}
    +
    \epsilon_G^{\mathrm{grid}}
    +
    \epsilon_G^{\mathrm{deriv}},
\]
with the production Matsubara order bracketed by neighbouring truncations, multiple stored-time grids, and independent derivative estimators. Across the primary domain the maximum relative $\epsilon_G$ is $0.0690\%$. Sampled ideal gaps are computed analytically on the Grover iteration grid, while the sampled-response uncertainty is propagated from independent truncation, grid, and ODE-solver checks of $D_{q,j}^{\mathrm{lin}}$. The nearest-threshold primary point,
$(\theta,\lambda,q)=(0.35,0.030,z)$, has
\[
    \eta_{\mathrm{step}}^{(1)}=0.640713,
    \qquad
    \eta_{\mathrm{step}}^{(1)}\in[0.640423,0.641003],
\]
so the deterministic numerical interval remains well separated from the transition at one; the upper uncertainty bound remains below one throughout the 108-point production domain.

The independently fitted finite-$\lambda$ comparator $\widehat G_q^{\mathrm{traj}}$ is numerically stable across the 18 timing-slope fits: the largest relative slope standard error is $0.1027\%$, and the largest leave-one-$\lambda$-out or reduced-range slope change is $0.1519\%$. These validation checks are not folded into the a priori $\epsilon_G$ budget.

For fixed response sign and a positive active sampled differential, the same quantities give the local relative-sensitivity estimates
\[
    \frac{|\mathrm d\eta_{\mathrm{cell}}|}{\eta_{\mathrm{cell}}}
    \lesssim
    \frac{|\mathrm dG_q|}{|G_q|}
    +
    \frac{|\mathrm d m_{0,q}^{\mathrm{dir}}|}{m_{0,q}^{\mathrm{dir}}},
    \qquad
    \frac{|\mathrm d r_{q,j}|}{r_{q,j}}
    \lesssim
    \frac{|\mathrm dD_{q,j}|}{D_{q,j}}
    +
    \frac{|\mathrm d\Delta_{0,j}|}{\Delta_{0,j}},
    \quad
    r_{q,j}=\frac{\lambda D_{q,j}}{\Delta_{0,j}}.
\]
Thus small stroboscopic margins or small sampled gaps amplify otherwise modest input uncertainty.

\subsection{Bath-robustness protocols}
\label{app:physical-input-robustness}

Numerical uncertainty and physical-input sensitivity are treated separately. The first bath study tests recalibrated threshold topology. It retains the C1--C4 families defined in Section~\ref{subsec:threshold-crossing-results} and recalibrates the first-order threshold independently at five bath settings,
\[
    (\gamma,T)\in
    \{(0.5,1.0),(0.4,1.0),(0.6,1.0),(0.5,0.8),(0.5,1.2)\},
\]
then evaluates full HEOM at
\[
    \lambda/\lambda_{\mathrm{step}}^{(1)}
    \in\{0.75,0.90,1.00,1.10,1.25\}.
\]
This gives 100 full-HEOM points for testing whether the crossing/non-crossing topology persists when the operating coordinate is recalibrated with the realized bath.

Table~\ref{tab:physical-input-threshold-robustness} records the family-level outcome and quantitative ranges for this topology test.

\begin{table}[h]
\caption{Recalibrated threshold-topology robustness of the four candidate families. For crossing families the reported interval is the five-setting HEOM crossing ratio $\lambda_{\mathrm{cross}}^{\mathrm{HEOM}}/\lambda_{\mathrm{step}}^{(1)}$; for controls it is the normalized HEOM sampled gap at the pre-specified endpoint $1.25\lambda_{\mathrm{step}}^{(1)}$.}
\label{tab:physical-input-threshold-robustness}
\small
\centering
\begin{tabular}{@{}lcccc@{}}
\toprule
Family & $(\theta,q)$ & Role & Five-setting outcome & Quantitative range \\
\midrule
C1 & $(0.35,z)$ & primary control & non-crossing $5/5$ & $0.198975$--$0.218677$ \\
C2 & $(0.35,x)$ & primary crossing & crossing $5/5$ & $1.132414$--$1.166880$ \\
C3 & $(0.15,z)$ & small-angle control & non-crossing $5/5$ & $0.034611$--$0.060104$ \\
C4 & $(0.125,x)$ & small-angle crossing & crossing $5/5$ & $1.019762$--$1.042908$ \\
\bottomrule
\end{tabular}
\end{table}

The second bath study tests fixed-operating-point mismatch. For each C1--C4 family, the physical coupling is fixed at
\[
    \lambda
    \in
    \{0.90,1.00,1.10\}\,
    \lambda_{\mathrm{step,baseline}}^{(1)},
\]
and the same five bath settings are applied without changing this physical $\lambda$. This gives 60 points. The fixed-baseline decision retains the baseline threshold classification, while the bath-recalibrated decision recomputes $\lambda_{\mathrm{step}}^{(1)}$ for the realized bath but leaves the physical operating coupling unchanged. A true bath-induced stopping change means that the full-HEOM winner at the perturbed bath differs from the baseline-bath winner for the same candidate and fixed operating ratio. These definitions are used for the exact-step, decision-change, and stopping-regret statistics in Section~\ref{subsec:physical-robustness-results}.

\subsection{Higher-order threshold diagnostics}
\label{app:higher-order-threshold}

The first-order screening criterion does not require an explicit second-order model, but the omitted terms are useful for diagnosing the finite-coupling validity boundary. If the local success trajectory admits
\[
    P_q(t;\lambda,\theta)
    =
    P_0(t;\theta)+\lambda\chi_q(t)+\lambda^2\zeta_q(t)+O(\lambda^3),
\]
write
\[
    t_q^*
    =
    t_0^*+\lambda G_q+\lambda^2 b_q+O(\lambda^3).
\]
Evaluating the stationarity condition at $t_0^*$ gives
\begin{equation}
    b_q
    =
    \frac{
      \tfrac12 P_0'''(t_0^*)G_q^2
      +\ddot\chi_q(t_0^*)G_q
      +\dot\zeta_q(t_0^*)
    }{\kappa_0}.
    \label{eq:second-order-peak-shift}
\end{equation}
The continuous-cell first-order approximation is therefore locally well separated from its second-order correction when
\begin{equation}
    |\lambda^2b_q|
    \ll
    m_{0,q}^{\mathrm{dir}}-\lambda|G_q|.
    \label{eq:second-order-cell-separation}
\end{equation}

For the sampled ordering, define the finite-coupling remainder relative to the first-order gap prediction by
\begin{equation}
    \mathcal E_q^{\mathrm{gap}}(\lambda)
    =
    \Delta_q^{\mathrm{HEOM}}(\lambda)
    -
    \left[\Delta_0-\lambda D_q\right].
    \label{eq:sampled-gap-remainder}
\end{equation}
Using the refinement protocol of Appendix~\ref{app:threshold-protocol}, the checks in Section~\ref{subsec:threshold-refinement-results} separate hierarchy truncation from the physical remainder in Eq.~\eqref{eq:sampled-gap-remainder}. The refined roots, root spans, and local remainder fits are reported there; the local fits are diagnostic only and are not used as a global second-order guarantee.

\section{Conditional Large-Instance Scaling}
\label{app:large-instance-details}

The ideal single-solution relation is
\[
    \theta(N)=\arcsin(N^{-1/2})\sim N^{-1/2}.
\]
Let $s_N=t_0^*(\theta(N))/\tau_G(\theta(N))$ be the continuous first-peak location in Grover-iteration units and define the normalized distance to the nearest half-integer cell boundary,
\[
    d_N=\operatorname{dist}\!\left(s_N,\mathbb Z+\frac12\right)
    =\frac{\min\{m_0^-,m_0^+\}}{\tau_G}.
\]
The non-resonant large-instance regime assumes that there exists $d_{\mathrm{nr}}>0$ such that $d_N\ge d_{\mathrm{nr}}$ for all sufficiently large instances in the chosen sequence. Since $\tau_G=4\theta/\Omega_G=\Theta(\theta)$ under fixed $\Omega_G$, the direction-aware margin is then $\Theta(\theta)$. If also $|G_q|=\Theta(\theta^\alpha)$, Eq.~\eqref{eq:formal-continuous-cell-threshold-scaling} follows immediately.

Without the non-resonance condition, $d_N$ can approach zero along subsequences and the relevant margin can be $o(\theta)$, so the scaling need not hold along such subsequences.

For cross-response timing, if $\lambda=O(\theta^{1-\alpha})$, $G_{xz}=\Theta(\theta^\alpha)$, and the second-order timing remainder remains uniformly $O(\theta)$, then $\Delta t_{xz}^*=O(\theta)$. Proposition~\ref{prop:quadratic-transfer-loss} then gives $R_{c\leftarrow d}=O(\theta^2)=O(N^{-1})$ for bounded local curvature. This yields the conditional large-instance transfer-loss scaling
summarized in the main text.


\section{Numerical Protocols and Convergence Diagnostics}
\label{app:numerical-protocols}

\subsection{Production HEOM convergence}

Hierarchy depth and Matsubara order are tested before the full-HEOM trajectories are used for quantitative analysis. Representative checks include ordinary weak coupling, the strong end of the primary domain, small-angle points, and a low-temperature regime. Successive Matsubara-order differences decrease through the production choice $N_k=6$, while the primary $\lambda\le0.03$ domain has only $O(10^{-6})$ trajectory and first-peak-time changes under the final $N_k=6$, $L=3\to4$ corner check. These tests motivate the production setting
\[
    L_{\mathrm{prod}}=3,
    \qquad
    N_{k,\mathrm{prod}}=6,
\]
without claiming exact infinite-order convergence. Production full-HEOM propagation uses QuTiP's \texttt{HEOMSolver} with the \texttt{dop853} integrator, relative tolerance $10^{-11}$, absolute tolerance $10^{-13}$, and an explicit fixed output-time grid; dense output is not requested, and no \texttt{max\_step}, \texttt{first\_step}, Jacobian, or stiffness-specific option is set explicitly. The linearized-response equations use \texttt{solve\_ivp}/DOP853 with relative tolerance $10^{-10}$, absolute tolerance $10^{-12}$, $t_{\max}=5$, and 2401 output points; sampled response values are obtained from the dense interpolant.

Because the transfer-loss extension reaches $\lambda=0.08$, the auxiliary strong-coupling endpoint at $\theta=0.50$, $\gamma=0.50$, and $T=1.00$ is tested separately for both coupling channels. At fixed $L=3$, the final $N_k=5\to6$ trajectory residuals are $2.36\times10^{-4}$ for $S_z$ and $2.21\times10^{-4}$ for $S_x$. The initial $L=3\to4$ trajectory residuals are larger, $1.14\times10^{-2}$ and $6.00\times10^{-3}$, so this auxiliary point is more hierarchy-depth sensitive than the primary domain. Extending the hierarchy to $L=6$ gives final $L=5\to6$ trajectory residuals $3.86\times10^{-4}$ for $S_z$ and $8.8\times10^{-5}$ for $S_x$, with first-peak-time residuals $5\times10^{-6}$ and $1.9\times10^{-5}$, respectively. The $L=6$ states satisfy the same trace, positivity, and Hermiticity screens. The sampled optimum is $j^*=1$ for both channels at every tested Matsubara order and hierarchy depth.

The full eight-point transfer grid is also recomputed at $L=6$ because the transfer-loss law depends on continuous peak times and cross-channel loss. The largest depth sensitivity occurs at $\lambda=0.08$; the quantitative changes and fitted-law comparison are reported in Section~\ref{subsec:transfer-loss-results}.

For the recalibrated bath-robustness scan, we also perform a targeted hierarchy-depth robustness check on C2 at $T=0.8$ and fixed $N_k=6$. Using the same $\lambda_{\mathrm{step}}^{(1)}=0.0783106488$ as the pre-specified robustness scan, the $1.10$ point ($\lambda=0.0861417$) has raw (unnormalized) sampled gap $+0.006887$ at $L=3$ and $+0.005815$ at $L=4$, with $j^*=2$ in both cases. The $1.25$ point ($\lambda=0.0978883$) has raw sampled gap $-0.014688$ at $L=3$ and $-0.016228$ at $L=4$, with $j^*=1$ in both cases. The gap magnitudes change by about $15.6\%$ and $10.5\%$, respectively, but the sign bracket and sampled optima are unchanged. Trace, positivity, Hermiticity, and first-peak-identifiability checks also pass.

For the fixed-operating-$\lambda$ $T+20\%$ cases, a separate paired depth check recomputes all 12 baseline and all 12 $T+20\%$ operating points at $L=4$ with $N_k=6$ and the physical couplings, candidate indices, and sample times held fixed. All $24/24$ stopping labels and all $24/24$ signed-gap signs agree with their $L=3$ references. At common $L=4$, the four $T+20\%$ bath-induced truth changes are the same four physical points as at $L=3$, and the same three points still change from a correct fixed-baseline decision to an incorrect recalibrated decision. The minimum absolute signed gap at $L=4$ is $2.91204\times10^{-4}$, and all 24 physicality checks pass. These checks vary hierarchy depth at fixed $N_k=6$ and support only targeted decision-label robustness, not full hierarchy convergence or Matsubara-order convergence.

The continuous first peak is extracted on the production trajectory by locating candidate maxima with \texttt{scipy.signal.find\_peaks} at prominence $0.001$, selecting the first returned candidate, and refining that grid maximum by a three-point quadratic interpolation when the local concavity, boundary, and refined-vertex bracket checks are satisfied. If no candidate is found, no peak is reported rather than substituting a different extremum. This continuous-peak rule is distinct from the discrete sampled-step argmax used for stopping decisions.

The production scan is retained only when the physical-state and first-peak extraction checks summarized in Table~\ref{tab:heom-reliability} are satisfied.

\begin{table}[h]
\caption{Reliability checks for the 108-point production HEOM scan.}
\label{tab:heom-reliability}
\small
\centering
\begin{tabular}{@{}lll@{}}
\toprule
Diagnostic & Pass criterion & Worst observed / result \\
\midrule
Trace error
& $<10^{-8}$
& $4.22\times10^{-15}$ \\
Minimum eigenvalue
& $\ge -10^{-8}$
& $-2.78\times10^{-17}$ \\
Hermiticity error
& $<10^{-8}$
& $0$ \\
First-peak extraction
& first detected peak; checked quadratic refinement
& $108/108$ pass \\
Completed production scan
& all requested points
& $108/108$ \\
\bottomrule
\end{tabular}
\end{table}
\FloatBarrier

\subsection{Production and coupling-hold-out protocol}
\label{app:production-protocol}

The 108-point production scan uses the operational three-point candidate set
$\mathcal J_3(\theta)$ from Section~\ref{subsec:grover-active-subspace}, with exact ties resolved by the smaller index for both prediction and full-HEOM observation. The same reliability screen is applied to every point, and no point is removed because it lies near a predicted threshold.

The coupling hold-out split is fixed before evaluation:
\[
    \lambda_{\mathrm{cal}}=\{0.005,0.010,0.015\},
    \qquad
    \lambda_{\mathrm{hold}}=\{0.020,0.025,0.030\}.
\]
This is a pre-specified hold-out in $\lambda$ within the same $(\theta,q,\gamma,T)$ grid, rather than an independent external parameter domain. The empirical timing comparator is fitted only on the calibration set, whereas the a priori linearized-HEOM response is reused without finite-$\lambda$ refitting. The hold-out result is reported in Section~\ref{subsec:broad-stability-results}.

\subsection{Threshold stress-test and refinement protocol}
\label{app:threshold-protocol}

The threshold study uses the four candidate families and five-point normalized coupling grid defined in Section~\ref{subsec:threshold-crossing-results}. The observed crossing families C2/C4 are refined by recomputing their sign-change brackets at $(L,N_k)=(3,6),(4,6),(3,8),(4,8)$ and applying bisection inside the original bracket; C1/C3 use the same hierarchy settings at the upper endpoint. The recalibrated topology scan updates $\lambda_{\mathrm{step}}^{(1)}$ at each perturbed $(\gamma,T)$ setting before reusing that normalized grid; the distinct fixed-operating-$\lambda$ mismatch protocol is defined in Appendix~\ref{app:physical-input-robustness}.

\subsection{Small-angle and Leng benchmark protocol}
\label{app:small-angle-protocol}

The small-angle benchmark is the 64-point Cartesian grid used in Section~\ref{subsec:broad-stability-results}. For the sampled-success objective, the HEOM ground truth is
\[
    j_{\mathrm{HEOM}}^P
    =
    \arg\max_{j\in\mathcal J_3}
    P_{\mathrm{HEOM}}(j\tau_G),
\]
the fixed ideal rule returns $j_0^*$, and the first-order rule returns $j_{q,\mathrm{pred}}^{*(1)}$.

The Leng benchmark uses the standard-$4\theta$ adaptation specified here rather than an ad hoc reconstruction from aggregate summary statistics. One complete Grover iteration is the effective iteration. Let $p_{\mathrm{rand}}=\sin^2\theta$ and let $P_0(j)$ denote the ideal standard-$4\theta$ sampled success probability. The adapted predicted-success curve is
\begin{equation}
    P_{\mathrm{Leng}}(j)
    =p_{\mathrm{rand}}
    +\bigl[P_0(j)-p_{\mathrm{rand}}\bigr]F_{\mathrm{iter}}^{\,j},
    \label{eq:leng-predicted-success}
\end{equation}
and the native restart-aware runtime objective and selection rule are
\begin{equation}
    r_{\mathrm{Leng}}(j)=\frac{j}{P_{\mathrm{Leng}}(j)},
    \qquad
    j_{\mathrm{Leng}}=\operatorname*{arg\,min}_{j\in\mathcal J_{\mathrm{Leng}}}r_{\mathrm{Leng}}(j),
    \label{eq:leng-runtime-objective}
\end{equation}
with deterministic smaller-$j$ tie breaking. For the small-angle HEOM benchmark, the standard-$4\theta$ grid supplies success indices $j\in\mathcal J_{\mathrm{grid}}(\theta)$; with $j_{\max}=\max\mathcal J_{\mathrm{grid}}(\theta)$, the runtime search uses $\mathcal J_{\mathrm{Leng}}=\{1,\ldots,j_{\max}\}$. The one-iteration fidelity is evaluated as
\[
    F_{\mathrm{iter}}=\frac12+\frac{\operatorname{Tr}A}{6},
\]
where $A$ is the Bloch transfer matrix of the noisy one-step channel in the ideal interaction frame. In the small-angle HEOM benchmark this quantity comes from the standard-$4\theta$ one-step reduced channel; in the circuit extension it is reconstructed from the exact incremental density-matrix channel between successive compiled Grover depths rather than from a gate-count proxy.

For the circuit extension, the 56 reconstructed fidelity values are finite and lie in $[0.684968073542,1]$. They decrease monotonically with device-noise scale for all eight targets. Table~\ref{tab:circuit-leng-fidelity} reports the scale-resolved mean and range; the complete pointwise values are included in the accompanying reproducibility artifact. All reconstructed incremental channels pass the trace-preservation, near-complete-positivity, and reconstruction-consistency checks.

\begin{table}[h]
\centering
\caption{Scale-resolved incremental-channel fidelity used by the circuit Leng predictor. Each row summarizes the eight marked targets at a fixed device-noise scale.}
\label{tab:circuit-leng-fidelity}
\begin{tabular}{@{}rrrr@{}}
\toprule
$s$ & Mean $F_{\mathrm{iter}}$ & Minimum & Maximum \\
\midrule
0.0 & 1.000000 & 1.000000 & 1.000000 \\
0.5 & 0.937176 & 0.936843 & 0.937510 \\
1.0 & 0.879022 & 0.878401 & 0.879644 \\
1.5 & 0.825182 & 0.824312 & 0.826052 \\
2.0 & 0.775326 & 0.774244 & 0.776408 \\
2.5 & 0.729152 & 0.727891 & 0.730415 \\
3.0 & 0.686381 & 0.684968 & 0.687796 \\
\bottomrule
\end{tabular}
\end{table}

As an implementation check, the adapted Leng rule reproduces the 64-point pre-specified small-angle native-runtime benchmark on $64/64$ points with zero runtime regret. Because no noise-induced Leng step change occurs anywhere on that broad small-angle grid, this $64/64$ result is used primarily as an implementation-consistency check rather than as evidence that distinguishes competing stopping methods. The sampled-success and restart-aware runtime objectives remain distinct throughout: $P_{\mathrm{Leng}}$ is used to select the Leng runtime step, while exact HEOM or exact-density success probabilities are used to evaluate success-argmax agreement and success regret. Finite-shot frequencies are not used in the Leng construction.

For the 56-point circuit extension, the Leng runtime search is restricted to the positive completed-iteration depths $\mathcal J_{\mathrm{Leng}}=\{1,2,3\}$. The zero-iteration circuit remains available to the exact sampled-success argmax on $\mathcal J_{\mathrm{circ}}=\{0,1,2,3\}$, but it is excluded from Eq.~\eqref{eq:leng-runtime-objective} because $j=0$ would give a trivial zero numerator rather than a completed Grover-query attempt. To compare both methods on the same trajectories, define the exact sampled-success optimum and regret by
\[
    j_{\mathrm{succ}}^*
    =\operatorname*{arg\,max}_{j\in\mathcal J_{\mathrm{circ}}}P_{\mathrm{circ}}(j),
    \qquad
    R_{\mathrm{succ}}^{(m)}
    =P_{\mathrm{circ}}(j_{\mathrm{succ}}^*)-P_{\mathrm{circ}}(j_m),
\]
and define the exact restart-aware runtime truth on completed iterations by
\[
    r_{\mathrm{exact}}(j)=\frac{j}{P_{\mathrm{circ}}(j)},
    \qquad
    j_{\mathrm{rt}}^*
    =\operatorname*{arg\,min}_{j\in\{1,2,3\}}r_{\mathrm{exact}}(j),
    \qquad
    R_{\mathrm{rt}}^{(m)}
    =r_{\mathrm{exact}}(j_m)-r_{\mathrm{exact}}(j_{\mathrm{rt}}^*).
\]
Here $m\in\{\text{first order},\text{Leng}\}$, with the same deterministic smaller-$j$ tie rule. Applying these definitions gives the same-domain circuit comparison reported in Section~\ref{subsec:circuit-validation}. Because $j_{\mathrm{rt}}^*=1$ at every circuit point, the Leng runtime result is interpreted as a consistency check on this grid rather than as evidence that the method discriminates among multiple runtime-optimal depths.

\subsection{Aggregate performance and screening protocol}
\label{app:screening-workflow-protocol}

The five HEOM validation datasets contain 346 dataset-instance rows but 260 unique physical points after cross-dataset deduplication. Dataset-specific rows are retained for provenance, while aggregate classification, regret, and physical solver-workload statistics use the 260-point domain. A point is positive when the full-HEOM sampled optimum differs from the ideal sampled optimum, $j_q^*\neq j_0^*$; otherwise it is negative. Exact-step accuracy is evaluated on the complete sampled index, and stopping regret is
\[
    R_{\mathrm{stop}}
    =
    P_{\mathrm{HEOM}}(j_q^*\tau_G)
    -
    P_{\mathrm{HEOM}}(j_{\mathrm{rule}}\tau_G),
\]
with round-off-scale negative values clipped to zero only within numerical tolerance.

For the screening analysis, the first-order sampled-gap indicator is evaluated before any finite-$\lambda$ HEOM decision is used. When a propagated numerical interval is available, its conservative upper edge is used for the screening comparison; otherwise, the nominal screening value is used. Across the 260 unique physical points, 96 use only interval-upper values, 152 use only nominal values, and 12 appear in duplicated dataset rows with both forms of screening information; none of these duplicate ranges straddles $\eta_{\mathrm{esc}}=0.70$, $0.75$, or $0.80$. The primary rule uses $\eta_{\mathrm{esc}}=0.75=1-\delta_{\mathrm{esc}}$ with $\delta_{\mathrm{esc}}=0.25$ as a pre-specified safety buffer below the nominal first-order transition at unity. This operational buffer is motivated by the separated-margin logic of Proposition~\ref{prop:first-order-sampled-cell}; it is not fitted to the validation outcomes. A point is accepted without escalation only if the conservative indicator is below $\eta_{\mathrm{esc}}$ and the first-order sampled rule preserves $j_0^*$; all other points are escalated to finite-$\lambda$ HEOM. The thresholds $0.70$ and $0.80$ are retained only as sensitivity checks. The validated broad sets terminate at a maximum conservative screening value of $0.6410034$, while the stress grids begin at the $0.75$ lower edge; no tested point lies in $(0.6410034,0.75)$. Because low-order acceptance is strict, points at exactly $0.75$ are escalated under the primary rule.

Solver-call accounting is reported after physical-point deduplication: the all-HEOM reference requires 260 full-HEOM stopping propagations, whereas the primary workflow uses 100 full-HEOM propagations plus 44 unique linearized-response solves. A same-environment benchmark on five representative validated configurations gives a pooled median linearized/full-HEOM runtime ratio $r=17.2396$. The resulting coarse workflow/full cost ratio is $3.3021$, with a range $3.2343$--$3.3021$ across three timing aggregations. Thus the current implementation reduces full-HEOM evaluations but not total wall-clock cost. A likely contributor is that the linearized path constructs a dense 2401-point response trajectory while the stopping HEOM path uses sparse sampled times. The timing result is specific to the recorded implementation and benchmark environment.

Dataset-level exact-step, truth-change, and escalation accounting is reported in Table~\ref{tab:screening-escalation} as overlapping dataset-instance provenance; the Fixed-Ideal exact count in each row equals the number of HEOM-unchanged instances in that dataset. Aggregate regret uses the 260 unique physical points: among the 15 genuine-change points, mean regret is $7.84063\times10^{-3}$ for Fixed Ideal and $0$ for First order; among the 245 stable points, the means are $0$ and $4.41833\times10^{-3}$, respectively. First order has 45 early calls, 215 exact calls, and zero late calls on the unique-point domain; every non-exact call is exactly one sampled step early, with mean early-call regret $0.02405536$.

On the 20 escalated points with available NZ2 trajectories, NZ2 is treated only as an optional comparator. First-order and NZ2 disagree on six points; all six are resolved in favor of NZ2 by the full-HEOM truth in this dataset. This empirical observation does not promote NZ2 to a general arbiter: any disagreement remains escalated to full HEOM.

\subsection{Same-bath NZ2 comparison protocol}
\label{app:nz2-protocol}

NZ2 uses the same $H_S$, $S_q$, Drude--Lorentz correlation function, and $(\lambda,\gamma,T)$ as HEOM, with no HEOM-derived damping, phase, or timing parameter. The threshold comparison reuses exactly the 20 C1--C4 candidate points from the pre-specified stress-test grid. Every NZ2 trajectory is screened for trace, Hermiticity, positivity, and first-peak identifiability before sampled stopping quantities are extracted. The same candidate set and tie rule are then applied to HEOM and NZ2. Crossing roots for C2 and C4 are refined within the corresponding pre-specified sign-change brackets, while C1/C3 are retained as non-crossing HEOM controls.

\subsection{Circuit protocol}
\label{app:circuit-protocol}

The circuit experiment uses a standard $n=3$ Grover circuit, all eight marked targets, and
\[
    \mathcal J_{\mathrm{circ}}=\{0,1,2,3\}.
\]
A seeded Qiskit \texttt{GenericBackendV2} fake-device configuration is fixed with three qubits, bidirectional line connectivity $0\leftrightarrow1\leftrightarrow2$, native basis gates \texttt{rz}, \texttt{sx}, \texttt{x}, and \texttt{cx}, backend seed 20261007, transpiler seed 1234, optimization level 0, and initial layout $[0,1,2]$. The compiled circuits are generated once and reused for all noise-scale values. The seeded calibration has qubit-dependent $T_1$ and $T_2$, direction-dependent two-qubit errors and gate durations, and nonzero measurement error. The scale $s$ multiplies the backend-reported gate errors and gate durations; $s=1$ is the nominal seeded fake-device calibration and larger $s$ values are controlled stress tests.

The circuit-native response is computed only from exact density-matrix simulations as
\begin{equation}
    \chi_{\mathrm{circ}}(j)
    =
    \frac{P_{\mathrm{circ}}(j;s_\epsilon)-P_0(j)}{s_\epsilon},
    \qquad
    s_\epsilon=0.005,
    \label{eq:circuit-native-response}
\end{equation}
with $s_\epsilon=0.010$ retained as an independent finite-difference consistency check. The exact response signs agree on all 32 target--depth points across the two perturbation scales. Finite-shot frequencies are not used to define or estimate $\chi_{\mathrm{circ}}$.

The exact stopping scan uses seven noise scales $s\in\{0,0.5,1,1.5,2,2.5,3\}$ and excludes readout error to isolate sampled ordering. The finite-shot layer uses the same 56 target--scale groups, 4096 shots per stopping decision, and 100 deterministic seeds per group, for 5600 seed-level decisions in total. Raw readout error is retained and no readout mitigation is applied. Exact ties choose the smaller $j$. Finite-shot sampling and unmitigated readout error jointly affect the selected stopping index, and their individual contributions are not separated; regret is nevertheless scored against the exact density-matrix probabilities at the exact and selected indices. Across the full range, gap/failure correlations are Pearson $-0.532110$ and Spearman $-0.753107$, with 204 one-step-early versus 10 one-step-late errors. At the hard boundary $s=1.5$, all eight exact winners are $j=2$ over runner-up $j=1$, with top-two gaps from $0.004584781$ to $0.006980508$ (median $0.005780852$). The finite-shot layer therefore tests sampling of the stopping decision, not estimation of the response coefficient in Eq.~\eqref{eq:circuit-native-response}.

For the circuit Leng comparison, $F_{\mathrm{iter}}$ is reconstructed from the exact incremental channel associated with one complete Grover iteration, and finite-shot data are excluded from both the fidelity and predictor. The Leng runtime rule is evaluated on $j\in\{1,2,3\}$, while sampled-success truth is evaluated from the same exact-density probabilities on $\mathcal J_{\mathrm{circ}}$. The objective definitions and fidelity values are given in Appendix~\ref{app:small-angle-protocol}; the corresponding numerical comparison is reported in Section~\ref{subsec:circuit-validation}.


\end{document}